\documentclass{article}

\usepackage[round]{natbib}

\usepackage{multiequations}
\usepackage{graphicx}
\usepackage{xcolor}
\usepackage{amsfonts}
\usepackage{amsmath}
\usepackage{amssymb}
\usepackage{empheq} 
\usepackage[hidelinks]{hyperref}
\usepackage{enumerate}  
\usepackage[loadonly]{enumitem}  
\usepackage{bbm}

\usepackage{geometry}

\newcommand*{\be}{\begin{equation}}
\newcommand*{\ee}{\end{equation}}
\newcommand*{\bea}{\begin{eqnarray}}
\newcommand*{\eea}{\end{eqnarray}}
\newcommand*{\bal}{\begin{align}}
\newcommand*{\eal}{\end{align}}
\newcommand*{\bme}{\begin{multiequations}}
\newcommand*{\eme}{\end{multiequations}}

\newcommand{\defi}{\stackrel{\triangle}{=}}
\newcommand{\dv}{\bnabla\! \bcdot\!}

\newcommand{\id}{\mbs{\mathbb{I}}_d}
\providecommand\bnabla{\boldsymbol{\nabla}}
\providecommand\bcdot{\boldsymbol{\cdot}}

\newcommand\Ros{\mbox{\textit{R\scriptsize o}}}  
\newcommand\Bu{\mbox{\textit{B\scriptsize u}}}  
\newcommand\Fr{\mbox{\textit{F\scriptsize r}}}  
\newcommand\Eu{\mbox{\textit{E\scriptsize u}}}  

\renewcommand*{\Omega}{\varOmega}
\renewcommand*{\Sigma}{\varSigma}

\newsavebox{\astrutbox}
\sbox{\astrutbox}{\rule[-5pt]{0pt}{20pt}}

\newcommand{\R}{\mathbb{R}}

\def\squarebox#1{\hbox to #1{\hfill\vbox to #1{\vfill}}}

\newcommand{\Dt}{\mathbb D}

\newcommand{\w}{\boldsymbol{w}}

\newcommand{\B}{\boldsymbol{B}}
\newcommand{\bsigma}{\boldsymbol{\sigma}}

\newcommand{\xx}{\boldsymbol{x}}
\newcommand{\kk}{\boldsymbol{k}}

\newcommand{\zz}{\boldsymbol{z}}

\newcommand{\nab}{\boldsymbol{\nabla}}

\newcommand{\transp}{^{\scriptscriptstyle T}}
\newcommand{\tr}{\mathrm{tr}}

\newcommand{\Exp}{\mathbb{E}}

\newcommand{\dif}{{\mathrm{d}}}
\newcommand{\mbs}[1]{\ensuremath{\boldsymbol{#1}}}

\newcommand{\dBt}{\dif  \boldsymbol{B}_t}

\begin{document}

%

\title{Geophysical flows under location uncertainty, Part II\\
Quasi-geostrophy and efficient ensemble spreading}


\author{V. Resseguier${\dag\ddag}$$^{\ast}$\thanks{$^\ast$Corresponding author. Email: valentin.resseguier@inria.fr
\vspace{6pt}},
 E. M\'emin${\dag}$
and B. Chapron${\ddag}$
\\\vspace{6pt}  ${\dag}
$Inria, Fluminance group, Campus universitaire de Beaulieu, Rennes Cedex 35042, France\\ ${\ddag}
$Ifremer, LOPS, Pointe du Diable, Plouzan\'e 29280, France
}

\maketitle

\begin{abstract}
Models under location uncertainty are derived assuming that a component of the velocity is uncorrelated in time. The material derivative is accordingly modified to include an advection correction, inhomogeneous and anisotropic diffusion terms and a multiplicative noise contribution. In this paper, simplified geophysical dynamics are derived from a Boussinesq model under location uncertainty. Invoking usual scaling approximations and a moderate influence of the subgrid terms, stochastic formulations are obtained for the stratified Quasi-Geostrophy (QG) and the Surface Quasi-Geostrophy (SQG) models. Based on numerical simulations, benefits of the proposed stochastic formalism are demonstrated. A single realization of models under location uncertainty can restore small-scale structures. An ensemble of realizations further helps to assess model error prediction and outperforms perturbed deterministic models by one order of magnitude. Such a high uncertainty quantification skill is of primary interests for assimilation ensemble methods. MATLAB\textsuperscript{\sffamily\textregistered} code examples are available online.


\end{abstract}

\section{Introduction}

Ensemble forecasting and filtering are widely used in geophysical sciences for forecasting and climate projection. In practice, dynamical models are randomized through their initial conditions and a Gaussian error model, and are generally found to be underdispersive \citep{mitchell2012data,Gottwald-Harlim13,berner2011model,Snyder15} with a low variance. As a consequence, errors are underestimated and observations are hardly taken into account. Corrections are considered by incorporating inflation procedures or hyperprior to increase the variance of ensemble Kalman filters \citep{Anderson99, Bocquet15}. However, such corrections do not provide an accurate spatial localization of the errors. 

Another difficulty of  ensemble methods lies in the huge dimensions of the involved state spaces. For obvious computational reasons, ensembles for geophysical applications appear constrained and limited to small sizes. It thus becomes primordial to build strategies to best track the most likely dynamical events. From this point of view, ensemble simulations and stochastic dynamics have clear advantages over the deterministic models. 

The simplest random models are defined from Langevin equations with linear damping and additive isotropic Gaussian noise, as, for instance, the linear inverse models \citep[e.g.][]{penland1994balance,penland1995optimal}, or the Eddy-Damped Quasi Normal Markovian (EDQNM) models \citep[e.g.][]{orszag1970analytical,Leith71,chasnov1991simulation}. Among other empirical stochastic models, the Stochastic Kinetic Energy Backscatter (SKEBS) \citep{Shutts05,berner2009spectral,berner2011model} and the Stochastic Perturbed Physics Tendency scheme (SPPT) \citep{Buizza99} introduce correlated multiplicative noises. SPPT and SKEBS methods have been successfully applied in operational weather forecast centers \citep{Franzke15}. To target highly non-Gaussian distribution of fluid dynamics properties, an attractive path is to infer randomness from physics \citep{Berner15}. For this purpose, the time-scale separation assumption is convenient. \cite{Hasselmann76} already relied on it for geophysical fluid dynamics. This assumption is the foundation of averaging and homogenization theories \citep{kurtz1973limit,papanicolaou1974asymptotic,givon2004extracting,Gottwald13,mitchell2012data,Gottwald-Harlim13,Franzke15,gottwald2015stochastic}.
A successful application of homogenization theory in geophysics is the MTV algorithms \citep{Majda99,Majda01,Franzke05,Majda08}. 
The homogenized dynamics is cubic with correlated additive and multiplicative (CAM) noises. This noise structure is able to produce intermittency and extreme events. In practice, the non-linearity of the small-scale equation (fast dynamics) is conveniently replaced by a noise and a damping terms before the homogenization procedure. Noise statistics are estimated from data, with Gaussian assumptions.

In \cite{resseguier2016geo1}, following \cite{Memin14}, another approach has been considered to help derive models under location uncertainty based on stochastic calculus and the Ito-Wentzell formula \citep{Kunita}. \cite{mikulevicius2004stochastic} and \cite{flandoli2011interaction} already introduce this methodology. Yet, their works mostly focused on pure mathematical aims: existence and uniqueness of SPDE solutions. For our more practical purpose, the large-scale is understood as sub-sampled in time, and the remaining small-scale velocity component is then considered as uncorrelated in time.

Starting with the definition of the revised transport under location uncertainty (section 2), developments are then carried out to derive and analyze the stochastic versions of Quasi-Geostrophy (QG) and Surface Quasi-Geostrophy (SQG) models with a moderate influence of sub-grid terms (section 3). Numerical results highlight the potential of these models under location uncertainty, especially for ensemble forecast (Section 4).

\section{Models under location uncertainty}
\label{Models under location uncertainty}
This section briefly outlines main theoretical results discussed in \cite{resseguier2016geo1}. The velocity is decomposed between a possibly random large-scale component, $\w$, and a time-uncorrelated component, $\bsigma \dot{\B}$. The latter is Gaussian, correlated in space with possible inhomogeneities and anisotropy. Hereafter, this unresolved velocity component will further be assumed to be solenoidal. To parameterize those spatial correlations, we apply an infinite-dimensional linear operator, $\bsigma$, to a $d$-dimensional space-time white noise\footnote{Formally each coefficient of $(t\mapsto \B_t)$ is a cylindrical $I_d$-Wiener process (see \cite{DaPrato} and \cite{Prevot07} for more information on infinite dimensional Wiener processes and cylindrical $I_d$-Wiener processes).}, $\dot{\B}$.

In time, the velocity is irregular. The material derivative, $D_t$, is then changed. In most cases, it coincides with the stochastic transport operator, $\Dt_t$, defined for every field, $\Theta$, as follows:
\bea
{ \Dt}_t \Theta \ 
&\defi&
\underbrace{
\dif_t \Theta
}_{
\substack{
 \defi \  \Theta(\xx,t+\dif t) - \Theta(\xx,t) \\
\text{Time increment}
}
}
+ 
\underbrace{
\left ({\w}^\star\dif t + \bsigma \dif\B_t \right)\bcdot \nab \Theta
}_{\text{Advection}}
 - 
\underbrace{
\nab\bcdot \left ( \frac{1}{2} \mbs a  \nab \Theta \right )
}_{\text{Diffusion}}
\dif t ,
\label{Mder}
\eea 
where the time increment term $\dif_t \Theta$ is used in place of the partial time derivative, as $\Theta$ is in general non-differentiable. The diffusion coefficient matrix, $\mbs a$, is solely defined by the one-point one-time covariance of the unresolved displacement per unit of time:
\bea 
\mbs a = 
\bsigma \bsigma \transp =
\frac{
\Exp \left \{ \bsigma \dBt \left( \bsigma \dBt  \right)\transp  \right \} 
}{\dif t}
,
\label{balance}
\eea
and the modified drift is given by
\bea
\w^\star = \w   - \frac{1}{2} ( \nab\bcdot \mbs a)\transp
.
\eea
With this modified material derivative \eqref{Mder}, the transport equations under location uncertainty involve three new terms: a modification of the large-scale advection ($\w^\star$ instead of $\w$), an inhomogeneous and anisotropic diffusion and a multiplicative noise. This random forcing is directly related to the advection by the unresolved velocity. 

 For incompressible flows ($\dv \w^\star=0$), the energy of any tracer, $\Theta$, is conserved for each realization:
 \bea
 \dif \int_\Omega \Theta^2
 =0
 ,
 \eea
 where $\Omega$ is the spatial domain. This still holds for active tracers. The diffusion dissipates as much energy as the multiplicative noise is injecting it in the system. In particular, the (ensemble) mean of the energy, $\Exp \int_\Omega \Theta^2$, is conserved. This results ensures a constant balance between the energy of the mean and the (ensemble) variance. The energy fluxes in these stochastic models are more thoroughly described in \cite{resseguier2016geo1}.
 \\

 A random version of the Reynolds transport theorem can further be derived \citep{Memin14,resseguier2016geo1}. From this theorem, usual conservation of mechanics (mass, linear momentum, energy and amount of substance) can be expressed in a stochastic sense. Random Navier-Stokes and Boussinesq models can then be derived. This last model describes the stochastic transports of velocity and density anomaly, as well as incompressibility conditions.


\section{Mesoscales under moderate uncertainty}
\label{subsection Simplified stochastic oceanic models}


To simplify the stochastic Boussinesq model of \cite{resseguier2016geo1},
Quasi-Geostrophic (QG) models are developed for large horizontal length scales, $L$,  such as:
\begin{eqnarray}
\frac 1 {\Bu} 
= \left(  \frac {\Fr}{\Ros} \right)^2
= \left ( \frac L {L_d} \right )^2 
\sim 1
\text{ and }
 \frac 1 \Ros = \frac{L f_0} U \gg 1,
\end{eqnarray}
where $U$ is the horizontal velocity scale, $L_d \defi \frac{Nh}{f}$ is the Rossby deformation radius, $N$ is the stratification (Brunt-V\"ais\"al\"a frequency) and $h$ is the characteristic vertical length scale. The Rossby deformation radius explicitly defines the mesoscale range, over which both kinetic and buoyancy effects are important, and strongly interact. In the following, both differential operators Del, $\nab$, and Laplacian, $\Delta$, represent $2$D operators.


\subsection{Specific scaling assumptions}
Hereafter, we explicit scaling assumptions to derive the non-dimensional version of the stochastic Boussinesq model.

\subsubsection{Quadratic variation scaling}
\label{Scaling}

 Besides traditional ones, another dimensionless number, $\Upsilon$, is introduced to relate the large-scale kinetic energy to the energy dissipation due to the horizontal small-scale random component. In the following, $\bsigma_{H \bullet} $ stands for the horizontal component of $\bsigma$, $\mbs a_H$ for $\bsigma_{H \bullet} 
 \bsigma_{H \bullet}  \transp$ and $A_u$ for its scaling. The new dimensionless number is defined by:
\begin{equation}
\Upsilon
\defi  \frac{UL}{A_u }
= \frac{U^2}{A_u / T}.
\end{equation}
This number compares horizontal advective and diffusive terms in the momentum and buoyancy equations. 
This number can also be related to the ratio between the Mean Kinetic Energy (MKE), $U^2$, and the Turbulent Kinetic Energy (TKE), $A_u/T_{\sigma}$, where $T_\sigma$ is the small-scale correlation time.  This reads:
\bea
\Upsilon
= \frac{1}{\epsilon}\frac{MKE}{TKE} 
,
\eea
where $\epsilon=T_\sigma/T$ is the ratio of the small-scale to  the large-scale correlation times. This parameter, $\epsilon$, is central in homogenization and averaging methods \citep{Majda99,givon2004extracting,Gottwald13}. The number $\Ros / \Upsilon$ can then be stated to measure the ratio between sub-grid terms and the Coriolis force. In the usual deterministic case and the limit of small Rossby number, the predominant terms of the horizontal momentum equation then correspond to the geostrophic balance. In the stochastic case, this balance also applies from weak ($\Upsilon\gg1$) to moderate ($\Upsilon \sim 1$) uncertainty. However, if $\Upsilon/ \Ros$ is close enough to $O(1)$, this geostrophic balance is modified due to the diffusion effects introduced by the small-scale random velocity. Hereafter, developments focus on the moderate uncertainty case. \cite{resseguier2016geo3} deals with the strong uncertainty case.

To evaluate $\Upsilon$ for a given flow at a given scale, eddy viscosity or diffusivity values help the determination of $A_u$. \cite{boccaletti2007mixed} give some examples of canonical values. Then, the typical resolved velocity and length scale lead to $\Upsilon$. If no canonical values are known, absolute diffusivity or similar mixing diagnoses could be measured \citep{keating2011diagnosing} as a proxy of the variance tensor.

\subsubsection{Vertical unresolved velocity}
\label{vertical velocity Scaling}

The scaling to compare vertical to horizontal unresolved velocities is also considered:
\begin{eqnarray}
\label{scaling_vertical_sigma_dBt}
\frac { (\bsigma \dif \B_t)_{z} }{\| (\bsigma \dif \B_t)_{H}\| } \sim \frac{\Ros}{\Bu} D,
\end{eqnarray}
where $D=\frac h L$ is the aspect ratio and the subscript $H$ indicates horizontal coordinates. This scaling can be derived from the $\omega$-equation \citep{giordani2006advanced}.
For any velocity $\mbs{\mathfrak{u}} = (\mbs{\mathfrak{u}}_H,\mathfrak{w})\transp$, which scales as $(\mathfrak{U},\mathfrak{U},\mathfrak{W})\transp$, this equation reads
\begin{equation}
f_0^2 \partial_z^2 \mathfrak{w}
+ N^2 \Delta \mathfrak{w}
=\nab \bcdot \mbs Q
\approx
- \nab \bcdot \left (  \nab \mbs{\mathfrak{u}}_H \transp  \nab \mathfrak{b} \right)
\approx
- f_0\nab \bcdot  \left (  \nab  \mbs{\mathfrak{u}}_H  \transp \partial_z \mbs{\mathfrak{u}}_H^{\bot} \right),
\end{equation}
where  $\mathfrak{b}$ stands for the buoyancy variable and $\mbs Q$ for the so-called $\mbs Q$-vector. In its non-dimentional version, the $\omega$-equation reads:
\begin{equation}
\frac{\mathfrak{W}}{\mathfrak{U}}
\left (
\partial_z^2 \mathfrak{w}
+ \Bu \Delta \mathfrak{w}
\right)
\approx 
D \Ros
\nab \bcdot \mbs Q.
\end{equation}
At planetary scales, Burger number is small and the rotation dominates the stratification, $\frac{\mathfrak{W}}{\mathfrak{U}} \sim D \Ros 
$. At smaller scales, with  a larger Burger number,  the stratification dominates the rotation, $\frac{\mathfrak{W}}{\mathfrak{U}} \sim D  \Ros/\Bu $. For the small-scale velocity $\bsigma \dot{\mbs B}$, the latter is thus more relevant. 

Note that the angle between the small-scale component and the horizontal one can be assumed to be constrained by the angle between the isopycnical and the horizontal plane. Invoked to describe baroclinic instabilities theory, this statement helps to specify the anisotropy of the eddy diffusivity \citep{Vallis}. The argument of the orientation of the eddies activity with isentropic surfaces and the related mixing is also supported by several other authors \citep{Gent90,Pierrehumbert93}. 

In the case of QG models, the large and small Burger scaling cases lead to the same result: the unresolved velocity is mainly horizontal.
\begin{eqnarray}
\label{scaling_vertical_sigma_dBt_QGmoderate_uncertainty}
\frac { (\bsigma \dif \B_t)_{z} }{\| (\bsigma \dif \B_t)_{H}\| } \sim \frac{\Ros}{\Bu} D \ll D.
\end{eqnarray}
This is consistent with the assumption of a large stratification, {\em i.e.} flat isopycnicals, if we admit that the eddies activity appears preferentially along the isentropic surfaces. As a consequence, the terms $(\bsigma \dif \B_t)_z\partial_z$ scale as $\frac{\Ros}{\Bu} (\bsigma \dif \B_t)_H \bcdot \nab$. In the QG approximation, the scaling of the diffusion and effective advection terms including $\bsigma_{z \bullet}$ are one to two orders smaller (in power of $\Ros/\Bu $) than terms involving $\bsigma_{H\bullet}$. For any function $\xi$, the vertical diffusion $\partial_z( \frac{\bsigma_{z \bullet} \bsigma_{z \bullet} \transp} 2  \partial_z \xi)$ is one order smaller than the horizontal-vertical diffusion term $ \nab \bcdot \left( \frac{\mbs \bsigma_{H \bullet} \bsigma_{z \bullet} \transp} 2 \partial_z \xi \right)$ and two orders smaller than the horizontal diffusion term $ \nab \bcdot \left( \frac{\bsigma_{H \bullet} \bsigma_{H \bullet} \transp} 2 \nab \xi \right)$.

\subsubsection{Beta effect}
\label{beta effect Scaling}

At mid-latitudes, the related term, given by $\beta \defi \partial_y f $, is much smaller than the constant part of the Coriolis frequency. Nevertheless, it can govern a large part of the relative vorticity at large scales. The following scaling is thus chosen \citep{Vallis}:
\begin{eqnarray}
\beta y \sim \nab^{\bot} \bcdot \mbs u \sim \frac U L = \Ros f_0.
\end{eqnarray}

\subsection{Stratified Quasi-Geostrophic model under moderate uncertainty}
\label{subsubsection Stratified Quasi-Geostrophic model under moderate uncertainty}

The moderate uncertainty case corresponds to $\Upsilon \sim 1$. Horizontal advective terms and horizontal sub-grid terms are comparable.
 
 Following similar principles than those used to derive the deterministic stratified QG model \citep{Vallis}, a stochastic QG model can be derived (see Appendix \ref{appendix QG model under moderate uncertainty}). 
This QG solution corresponds to the limit of the Boussinesq solution when the Rossby number goes to zero.
The resulting potential vorticity (PV), $Q$, is then found to be conserved, along the horizontal random flow, up to three source terms:
\begin{multline}
\label{eq QG moderate uncertainty Dt}
\Dt^H_{t} Q
= 
  \frac{1}{2} \sum_{i,j\in H}  \partial_{ij}^2 \left( \nab^{\bot} a_{ij} \bcdot  \mbs u \right)     \dif t
-  \frac 1 2 \nab \bcdot \left( \nab \bcdot (\mbs a_H f) \right) \transp  \dif t
- \tr \left [ 
 \mbs S \mbs J
 \mbs S_{ \sigma \dif B_t}
    \right ]
   ,
\end{multline}
where the QG PV is:
\begin{eqnarray}
\label{QG PV moderate uncertainty}
Q \defi  \Delta \psi +f+ \left( \frac {f_0} N \right )^2 \partial_z^2 \psi,
\end{eqnarray}
$\psi$ is the streamfunction, $\mbs J 
= \begin{pmatrix} 0 & -1 \\ 1 & 0 \end{pmatrix}
$ is the $\frac \pi 2$ rotation matrix,
\bea 
\mbs S 
= \frac 12 \left [ \nab \mbs u \transp + \left( \nab \mbs u \transp \right )\transp \right ]
\text{ and }
\mbs S_{ \sigma \dif B_t} =
 \frac 12 \left [ \nab (\bsigma \dif \B_t)_H \transp + \left( \nab (\bsigma \dif \B_t)_H \transp \right )\transp \right ]
 \eea
 denotes the strain rate tensor of the horizontal resolved and unresolved velocities, $\mbs u$ and $ (\bsigma \dif \B_t)_H $, respectively. 
 To interpret the source terms, we rather focus on the material derivative of the PV: 
\bea 
\label{eq QG moderate uncertainty deriv material}
D^H_t Q
&= &
  \nab \bcdot ( \alpha \nab \psi )    \dif t
-  \frac 1 2 \nab \bcdot \left( \nab \bcdot (\mbs a_H f) \right) \transp  \dif t
- \tr \left [ 
 \mbs S \mbs J
 \mbs S_{ \sigma \dif B_t}
 \right ],
 \eea
with
\begin{eqnarray}
\alpha \transp
\defi 
\sum_i
\left( \nab \bsigma_{H_i}\transp \right)^2,
\end{eqnarray}
which can be decomposed into a symmetric part, positive or negative diffusion of the stream function, and an anti-symmetric part, skew diffusion advection of the stream function. Compared to the traditional QG model, this system includes two smooth (continuous) source/sink terms that depend on the variance tensor, and a random forcing term. The first source term in (\ref{eq QG moderate uncertainty deriv material}) is correlated in time and may decrease or increase the PV energy. This term is due to the spatial variations of both the diffusion coefficient and the drift correction. The second term takes into account interactions between the Coriolis frequency, including beta effects, and inhomogeneous sub-grid eddies. The last source term in (\ref{eq QG moderate uncertainty deriv material}) is a noise term, encoding the interactions between the resolved and the unresolved strain rate tensors. Uncorrelated in time, this noise increases the potential enstrophy along time. 

To further understand this source term, let us denote $\Xi$ and $\Lambda$ the eigenvalues associated with the stable directions ({\em i.e.} negative eigenvalue) of the strain rate tensors of the large-scale flow, $\mbs S$, and of the small-scale flow, $\mbs S_{ \sigma \dif B_t}$ respectively. We note $\theta$, the angle between these two stable directions
\bea
\label{QG moderate uncertainty angle stable direction}
- \tr \left [ 
 \mbs S \mbs J
 \mbs S_{ \sigma \dif B_t}
 \right ]
 &=&
 2 \underbrace{ \Xi \Lambda }_{>0}
 \sin \left( 2 \theta \right) .
\eea
The detailed derivation is provided in Appendix \ref{appendix QG model under moderate uncertainty}. This random source vanishes when the stable directions of $\mbs u$ and $(\bsigma \dif \B_t)_H$ are aligned or orthogonal. It is maximum and positive (respectively minimum and negative) when there is an angle of $\frac {\pi} 4$ (respectively $-\frac {\pi} 4$) between those directions. Around the local position $\xx$, stable and unstable directions of the large-scale velocity define 2 axes and 4 quadrants. As understood, the strain rate tensor does not depend on the local vorticity. Yet, an hyperbolic deformation will almost resemble a positive vorticity in the upper-left and bottom-right quadrants, and a negative vorticity in the upper-right and bottom-left quadrants. For $\theta = \frac {\pi} 4$, the stable direction of the small-scale velocity aligns along the upper-left to bottom-right direction. The small-scale velocity then compresses the flow in this direction and dilates the flow in the orthogonal direction (upper-right to bottom-left). The quadrants associated with a seemingly positive (resp. negative) vorticity are brought closer (resp. farther) to $\xx$. Accordingly, the vorticity increases at $\xx$. For $-\frac \pi 4$, the vorticity would decrease. 

Note the $\dBt$ factor has been omitted in the right-hand side of equation (\ref{QG moderate uncertainty angle stable direction}). This term remains a linear function of the uncorrelated noise $\zz \mapsto  \dBt(\zz)$. Whatever the angle between the stable directions, the source term always has a zero (ensemble) mean and increases the enstrophy since it is a term in $\dBt$. Equation (\ref{QG moderate uncertainty angle stable direction}) could then be used to define the horizontal inhomogeneous small-scale component of the velocity. If the conservation of PV is a strong constraint, this component can indeed be defined to ensure that its stable direction is always along or orthogonal to the stable direction of $\mbs u$.

A two-layer model could also be deduced from equation \eqref{eq QG moderate uncertainty Dt} or \eqref{eq QG moderate uncertainty deriv material}. This would help identifying the stochastic parameterization effects on the barotropic and baroclinic modes. In particular, the particular forms of the operator $\bsigma$ able to trigger barotropization effects can bemore efficiently studied.

In the stochastic QG model, the stream function $\psi$ is related to the buoyancy, $b$, the pressure, $p'$, and the velocity, $\mbs u$, by the usual relations:
\begin{eqnarray}
b= f_0 \partial_z \psi, \ 
 p' = \rho_b f_0 \psi 
 \text{ and } 
\mbs u = \nab^{\bot} \psi 
,
\end{eqnarray}
where $\rho_b$ is the mean (background) density.
The horizontal noise term, $(\bsigma \dif \B_t)_H$, appearing in both the horizontal stochastic material derivative  and in the $2\times2$ horizontal variance tensor, $\mbs a _H$, is in geostrophic balance with a pressure component uncorrelated in time. Due to their scaling, the vertical noise and its variance are neglected in the final equations.\\
For homogeneous turbulence conditions, the transport of PV \eqref{eq QG moderate uncertainty deriv material} simplifies. The variance tensor becomes constant, the first two source terms disappear, to give
\begin{eqnarray}
\label{eq QG moderate uncertainty deriv material homogeneous}
\Dt_{t}^H Q = D_t^H Q 
= - \tr \left [ 
 \mbs S \mbs J
 \mbs S_{ \sigma \dif B_t}
 \right ].
\end{eqnarray}
The transport of the PV (equation \eqref{eq QG moderate uncertainty Dt} or \eqref{eq QG moderate uncertainty deriv material homogeneous}) determines the dynamics of the fluid interior. Boundary conditions are then necessary to specify completely the dynamics. 


\subsection{Surface Quasi-Geostrophic model under moderate uncertainty}
\label{subsubsection Surface Quasi-Geostrophic model under moderate uncertainty}
A classical choice considers a vanishing solution in the deep ocean and a buoyancy transport at the surface \citep{Vallis,Lapeyre06}:
\bea
\label{BC of QG under moderate uncertainty}
\psi \underset{z \to - \infty}{\longrightarrow} 0
\text{ and }
D^H_t b_{|_{z=0}} = 
\Dt^H_{t} b_{|_{z=0}} =
0.
\eea

Assuming zero PV in the interior but keeping these boundary conditions leads to the Surface Quasi-Geostrophic model (SQG) \citep{blumen1978uniform,Held95,Lapeyre06,constantin1994formation,constantin1999front,constantin2012new}. Under the stochastic framework, the derivation is similar. The PV is indeed identical to the classical one (see equation (\ref{QG PV moderate uncertainty})), assuming zero PV in the interior and vanishing solution as $z \to -\infty$ unsurprisingly yields the same SQG relationship:
\bea
\label{SQG relation moderate uncertainty}
\hat b 
=
N \| \mbs k \|
\ 
\hat \psi .
\eea
The top boundary condition, equation \eqref{BC of QG under moderate uncertainty}, provides an evolution equation, namely the horizontal transport of surface buoyancy, in the stochastic sense:
\bea
\label{transport SQG moderate uncertainty}
\Dt_t^H b =  0.
\eea The time-uncorrelated component of the velocity, $\bsigma \dot{\mbs B}$, is divergence-free. Its inhomogeneous and anisotropic spatial covariance has then to be specified. The time-correlated component of the velocity is also divergence-free, with a stream function specified by the SQG relation \eqref{SQG relation moderate uncertainty}. The buoyancy is randomly advected, and the resulting smooth velocity component is random as well.\\

\subsection{Summary}


For simplified models, stochastic versions are derived for scaling assumptions related to the sub-grid terms. For moderate uncertainty, the PV is transported along the random flow up to three source terms. The first one, smooth in time, is due to spatial variations of the inhomogeneous diffusion and the drift correction. The second one, also smooth, encodes the interaction between inhomogeneous turbulence and Coriolis frequency. These terms disappear for an homogeneous turbulence. The last term, a time-uncorrelated multiplicative noise, involves the large-scale and the small-scale strain rate tensors. It is a source of potential enstrophy and its instantaneous value depends on the angle between the large-scale and small-scale stable directions. Assuming zero PV in the interior, a SQG model follows from this QG model.


\section{Numerical results}
\label{section Numerical results}

We focus on this $SQG_{MU}$ model (\ref{subsubsection Surface Quasi-Geostrophic model under moderate uncertainty}). 
A high-resolution deterministic SQG simulation provides a reference.
The MATLAB\textsuperscript{\sffamily\textregistered} codes are available online (\url{http://vressegu.github.io/sqgmu}). 
Numerical results are analyzed in terms of the resolution gains (when a single realization is simulated) and the potential for ensemble forecasting in estimating spatial and spectral reconstruction errors (for an ensemble of realizations).

 \subsection{Test flow}
 \label{Test flow}
 
\begin{figure}
\begin{center} 
    \textbf{High-resolution buoyancy}\par\medskip
\includegraphics[width=5cm]{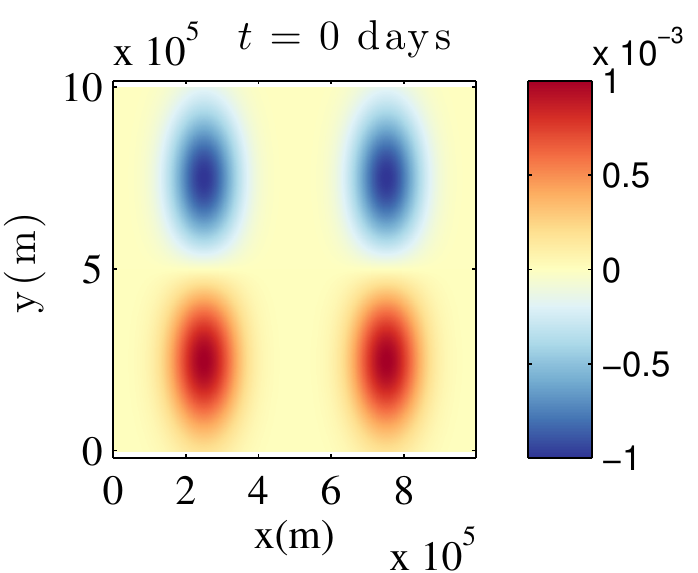}
\includegraphics[width=5cm]{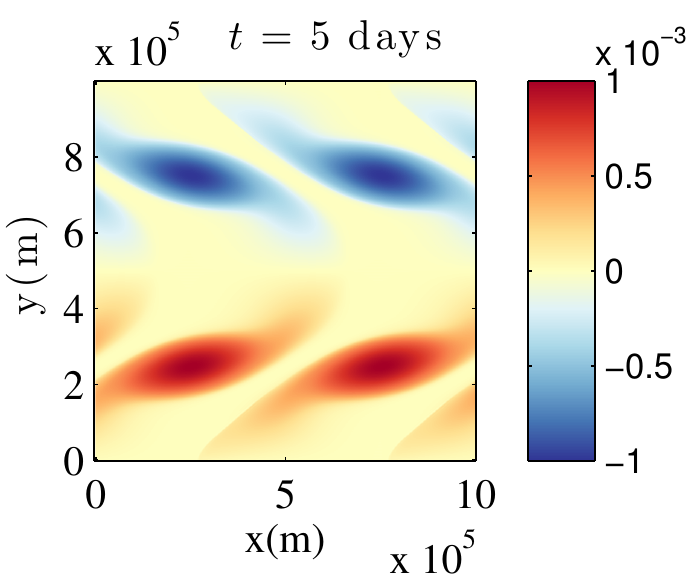}
\includegraphics[width=5cm]{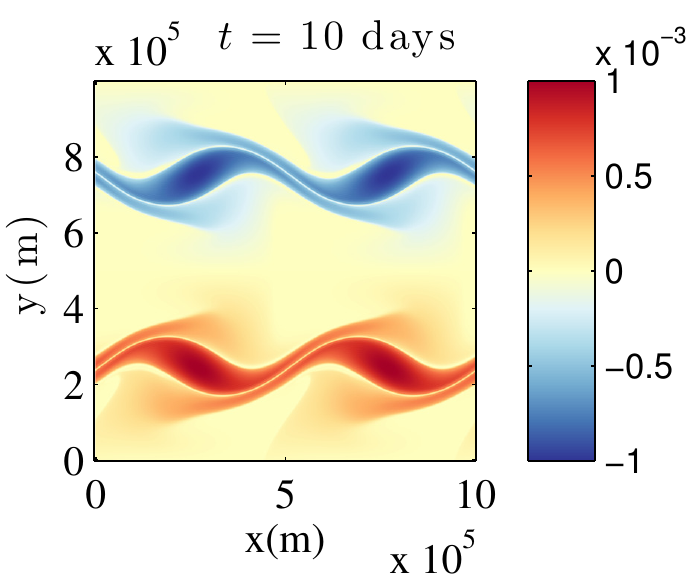}
\includegraphics[width=5cm]{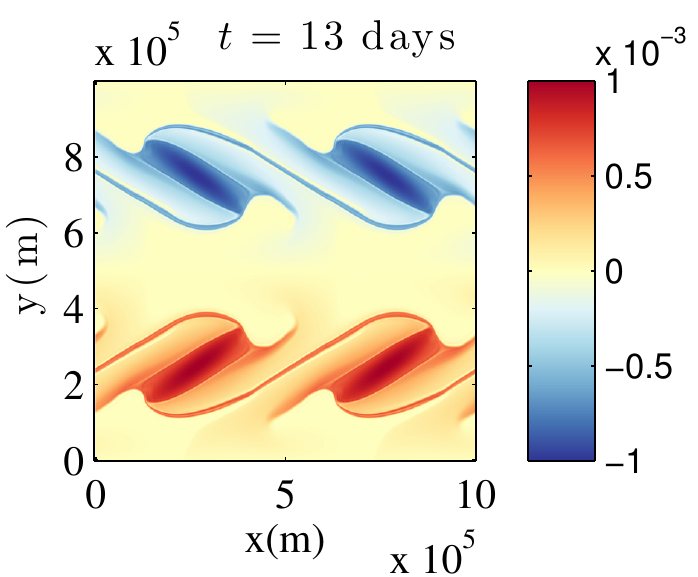}
\includegraphics[width=5cm]{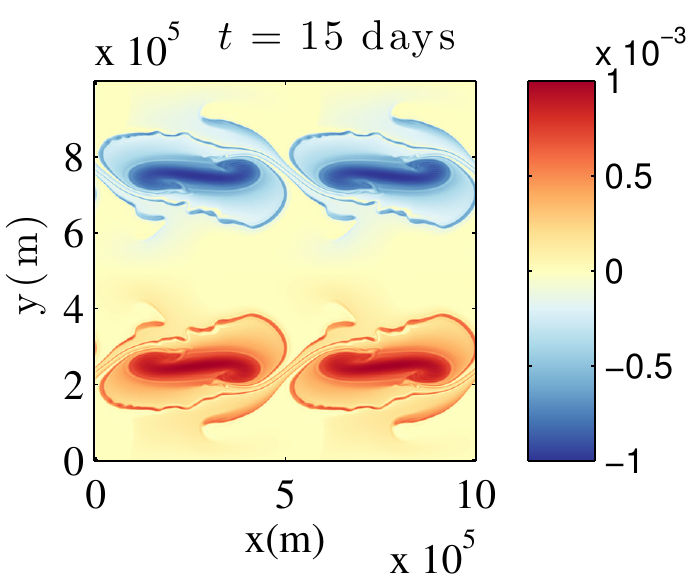}
\includegraphics[width=5cm]{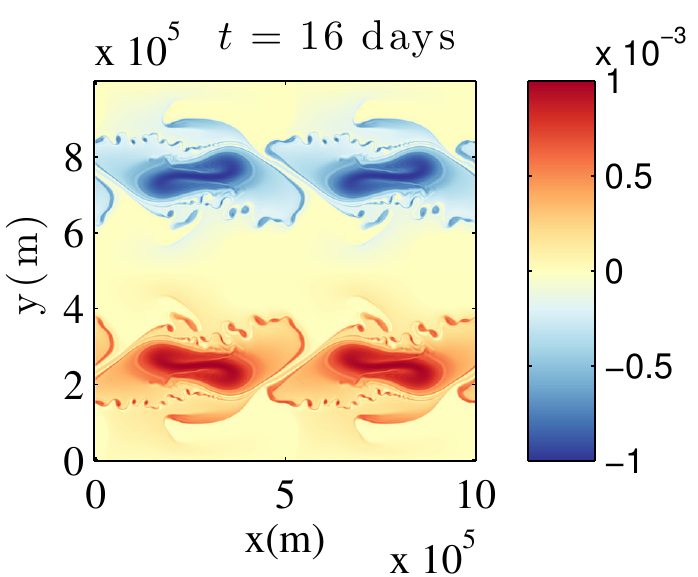}
\includegraphics[width=5cm]{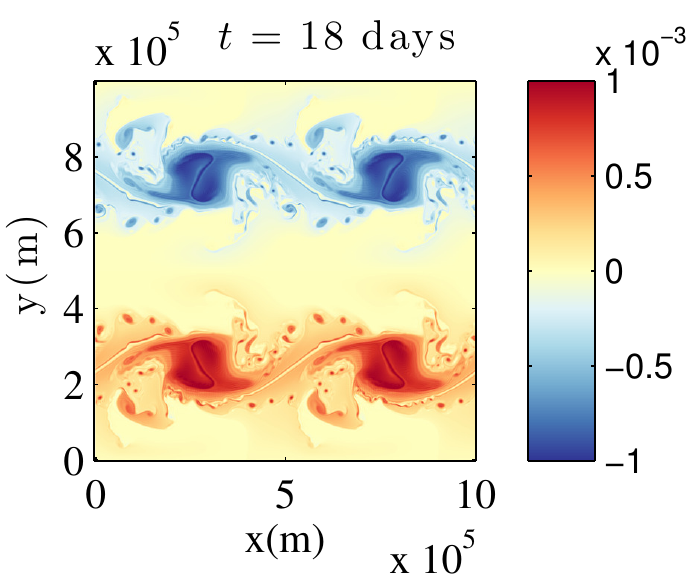}
\includegraphics[width=5cm]{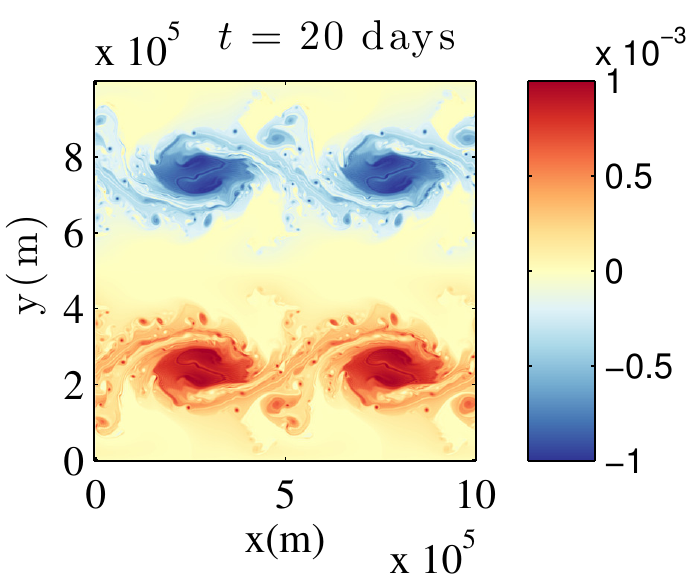}
\includegraphics[width=5cm]{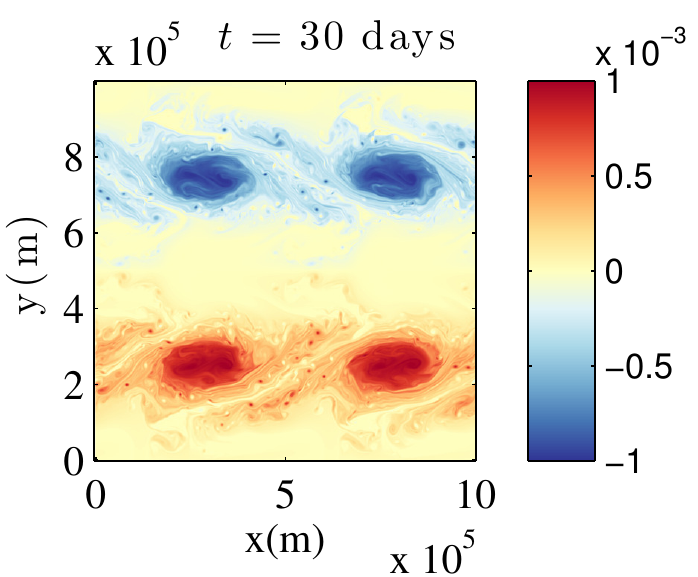}
\end{center}
\caption{Buoyancy ($m.s^{-2}$) at $t=0, 5, 10, 13, 15, 16, 20, 30$ days of advection for the usual SQG model at resolution $512^2$.}
\label{plot_HR_buoyancy}
\end{figure}

 The initial conditions defining the test flow, Figure \ref{plot_HR_buoyancy}, consist of a spatially smooth buoyancy field with two warm elliptical anticyclones and two cold elliptical cyclones given by:
 \bea 
 \label{initial condition SQG simu}
 b (\xx , t=0 ) 
 &=&
 F\left (\xx - 
 \begin{pmatrix}
250 \ \text{km} \\ 250 \ \text{km}
\end{pmatrix}
\right)
 + F\left (\xx - 
 \begin{pmatrix}
750 \ \text{km} \\ 250 \ \text{km}
\end{pmatrix}
\right) 
\nonumber\\
& &
- F\left (\xx - 
 \begin{pmatrix}
250 \ \text{km} \\ 750 \ \text{km}
\end{pmatrix}
\right)
- F\left (\xx - 
 \begin{pmatrix}
750 \ \text{km} \\ 750 \ \text{km}
\end{pmatrix}
\right),
 \eea
 with
 \bea 
  \label{initial condition SQG simu F}
F (\xx)
 & \defi &
 B_0 \exp \left( - \frac 12 \left(
 \frac {x^2} { \sigma_x^2}
 +
 \frac {y^2} { \sigma_y^2}
 \right) \right)
 \text{ and }
 \left\{
\begin{array}{r c l}
\sigma_x &=& 67 \ \text{km},\\
\sigma_y &=& 133 \ \text{km}.
\end{array}
\right.
 \eea 
 The size of the vortices is of order of the Rossby radius $L_d$. The buoyancy and the stratification have been set with: $B_0 = 10^{-3} m.s^{-2}$ and $N = 3 f_0$.
The Coriolis frequency is set to $1.028 \times 10^{-4} s^{-1}$ ($45^{\circ}$ N).
Periodic  boundaries conditions are considered.
 
 The deterministic high-resolution SQG reference model is associated with a spatial mesh grid of $512^2$ points, whereas the low-resolution (deterministic or stochastic) SQG models are run on $128^2$ points. The simulations have been performed through a pseudo-spectral code in space. As for the temporal discrete scheme the deterministic simulation relies on a fourth-order Runge-Kutta scheme, whereas the stochastic ones are based on an Euler-MaruyamaEuler-Maruyama scheme \citep{Kloeden-Platen}. 
For our application, the weak precision of this scheme is balanced by the use of a small time step. In all the simulations (deterministic and random, high-resolution and low-resolution), a standard hyperviscosity model is used:
 \bea 
 \Dt_t b =\alpha^{hv} \Delta^4 b \; \dif t,
 \eea  
with a coefficient $\alpha^{hv}=  (5 \times 10^{29} m^8.s^{-1} ) M_x^{-8} $ where $M_x$ denotes the meshgrid size ({\em i.e.} $128$ or $512$).

Figure \ref{plot_HR_buoyancy} displays the high-resolution buoyancy field at $t=0, 5, 10, 13, 15, 16, 20$ and $30$ days. During the first ten days, the vortices turn with slight deformation. 
Vortices of the same sign have their tails that draw closer. This creates high shears around four saddle points located at $(x,y)=(0,250),(500,250),(0,750)$ and $(500,750)$ (in km). A strong non-linearity in the neighborhood of a saddle point has been identified to become a major source of instability
\citep{constantin1994formation,constantin1999front,constantin2012new}. 
In our case, this effect is weak but yields an effective creation of turbulence $10$ days later. Shears create long and fine filaments, wrapping around the vortices until the $15^{th}$ day. At this time, the filaments become unstable, break and a so-called ``pearl-necklace" appears, characteristic of the SQG model, days $17$-$18$ in the simulation. These small vortices are then ejected from their orbits. Between days $17^{th}$ and $25^{th}$, they interact with the large vortices, the filaments and other small vortices, to create a fully-developed SQG turbulence orbiting around the four large vortices.

\subsection{Simulation of the random velocity}
\label{Simulation of the random velocity component}
To simulate the ${\rm SQG_{MU}}$ model (\ref{SQG relation moderate uncertainty}-\ref{transport SQG moderate uncertainty}), the covariance of the unresolved velocity $\bsigma \dot{ \B}$ must be specified. 
As this unresolved velocity field is assumed divergence-free, we introduce the following stream function linear operator, $\breve{\psi}_\sigma$, and its kernel, $\breve{\psi}_\sigma$:
\bea
\bsigma_H ( \xx) \dBt 
=
 \nab^{\bot} {\psi}_\sigma( \xx) \dif B_t ,
=
 \int_{\Omega} \dif \zz \;  \nab^{\bot}_{\xx} \breve{\psi}_\sigma (\xx,\zz) \dif B_t(\zz)
 .
\label{model solenoidal pour simu sigma}
\eea
As such, a single cylindrical Wiener process, $B_t$, is sufficient to sample our Gaussian process. This is specific to two-dimensional domains. In $3$D, a vector of $3$ independent $\id$-cylindrical Wiener processes, and a projection operator on the divergence-free vector space or a curl must be considered to simulate an isotropic small-scale velocity \citep{Memin14}. For a divergent unresolved velocity, equation \eqref{model solenoidal pour simu sigma} can additionally involve the gradient of a random potential, $\nab \tilde{\psi}_\sigma \dif B_t$.

\begin{figure}
\begin{center} 
\includegraphics[width=15cm]{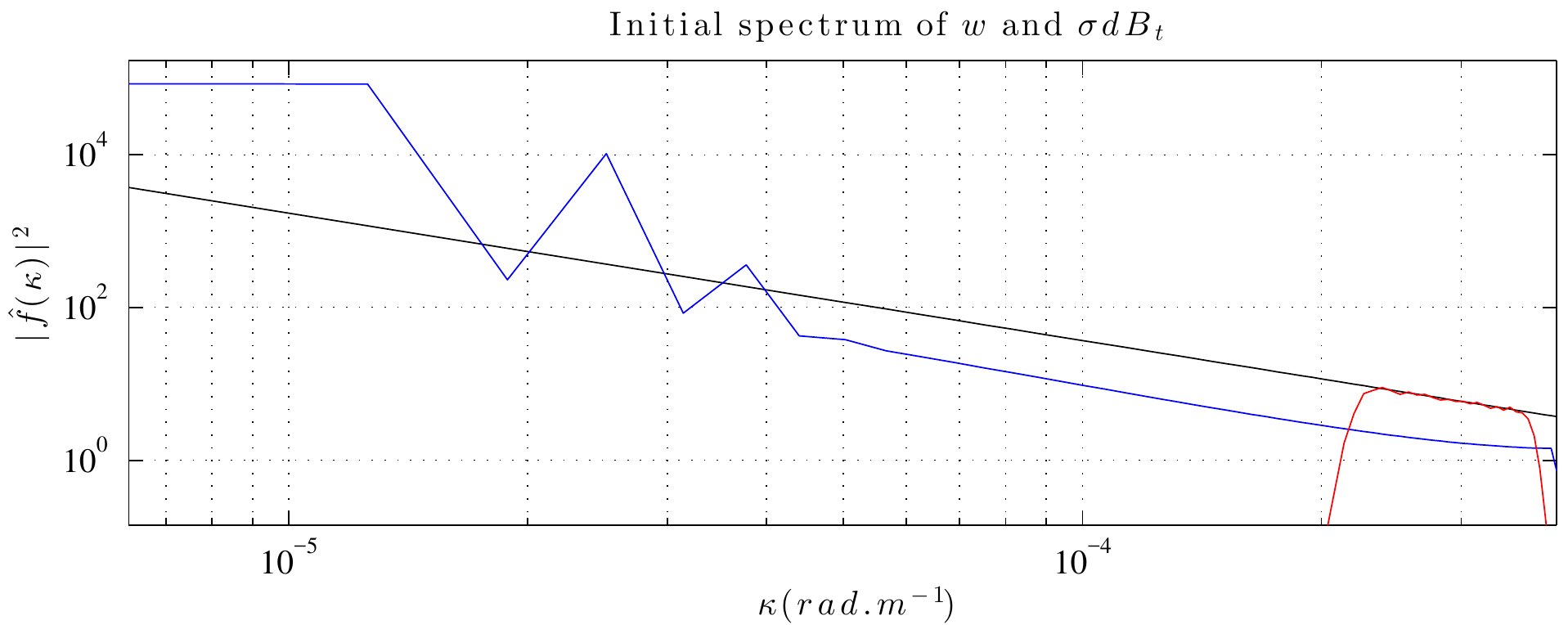}
\end{center}
\caption{Spectrum ($m^2.s^{-2}/($rad$.m^{-1})$) of $\mbs w$, at the initial time, in blue, spectrum of $\bsigma_H \dot{\mbs B}$ (up to a multiplicative constant), in red, and slope $- \frac 5 3$ in black. In the simulation performed, $\bsigma_H \dot{\mbs B}$ is restricted to a narrow spectral band. Thus, this velocity component almost only acts near the resolution cutoff, where the large-scale component, $\mbs w$, has a low energy.}
\label{spectrum_sigma}
\end{figure}

Then, similar to the  Kraichnan's model, a solenoidal homogeneous field can be considered:  \citep{Kraichnan68,kraichnan1994anomalous,Gawedzki95,Majda-Kramer}:
   \begin{eqnarray}
\bsigma_H( \xx) \dBt 
=
 \int_{\Omega} \dif \zz \;  \nab^{\bot}_{\xx} \breve{\psi}_\sigma (\xx-\zz) \dif B_t(\zz)
  =   \left( \nab^{\bot} \breve{\psi}_\sigma \star  \dif B_t \right)(\xx)
  .
 \end{eqnarray}
where $\star$ denotes a convolution. Although spatially inhomogeneous field would be more physically relevant, homogeneity greatly simplifies the random field simulation. Indeed, homogeneity in physical space implies independence between the Fourier modes
\bea
\widehat{\bsigma_H \dot{ \B}}(\mbs k)
=
i \kk^{\bot} \widehat{\breve{\psi}_\sigma }(\mbs k)\widehat{\dot{ B}}(\mbs k), 
\label{def sigma streamfunction fourier}
\eea
in the half-space $\mbs k \in
 (\R \times \R^{+*}) \cup (\R^{+} \times \{0\})
 $. Thus, the small-scale velocity can be conveniently specified from its omnidirectional spectrum:
\bea
\mbs k \mapsto
\frac 1 {\mu(\Omega)}
\Exp
\oint_{[0, 2 \pi]} \dif \theta_{\kk}  \| \kk\|
 \left \| \widehat{\bsigma_H \dot{ \B}}(\mbs k) \right \|^2
=
\frac {2 \pi } { \Delta t} \|\kk\|^3 \left | \widehat{\breve{\psi}_\sigma }\left(\| \mbs k \| \right) \right|^2 ,
\eea
where $ \mu(\Omega)$ is the surface of the spatial domain $\Omega$, $\theta_{\kk}$ is the angle of the wave-vector $\kk$ and $\Delta t$ the simulation time-step. Consistent with SQG turbulence, the omni-directional spectrum slope, denoted $s$, is fixed to $-\frac 5 3$. For $2$D Euler equations, the slope would be set to $-3$. If the small scales spectrum slope is unknown, the spectrum slope of the resolve scales -- estimated on line -- may enable to specify $s$ through a scale similarity assumption.
The unresolved velocity should be energetic only where the dynamics cannot be properly resolved.
Consequently, we apply to the spectrum a smooth band-pass filter, $f_{BP}$, which has non-zero values between two wavenumbers $\kappa_{min}$ and $\kappa_{max}$. The parameter $\kappa_{min}$ is inversely related to the spatial correlation length of the unresolved component. In practice, we set $\kappa_{max}$ to the theoretical resolution, $
\frac\pi{\Delta x}
$,
and $\kappa_{min}$ to the effective resolution (hereafter $\kappa_{min}=\kappa_{max}/2$). Figure \ref{spectrum_sigma} illustrates this spectrum specification.
The small scales' energy is specified by the diffusion coefficient $a_H$ and the simulation time step:
\bea
\label{link energy dissipation in homogeneous case}
\Exp \left( \bsigma_H \dot{ \B} \right ) \left( \bsigma_H \dot{ \B} \right )  \transp
= 
 \frac 1 { \Delta t} 
 \mbs a_H
=
 \frac 1 { \Delta t}
\begin{pmatrix}
a_H & 0  \\
0 & a_H
\end{pmatrix}.
\eea
The diagonal structure of the variance tensor is due both to incompressiblity and isotropy. The scalar variance tensor, $a_H$, is similar to an eddy viscosity coefficient. So, a typical value of eddy viscosity used in practice is a good proxy to setup this parameter. Otherwise, this parameter can be tuned. For this paper, it is set to $9 \; m^2.s^{-1}$. The time step depends itself, through the CFL conditions,
on both the spatial resolution and the maximum magnitude of the resolved velocity. Finally, equation \eqref{def sigma streamfunction fourier} writes:
  \begin{eqnarray}
  \label{numeric definition of sigma dBt}
\widehat{\bsigma_H \dot{ \B}}(\mbs k)
\defi
\frac A {\sqrt{\Delta t}} 
 \  i \mbs k^{\bot} 
f_{BP} \left(  \left \| \mbs k  \right \| \right ) 
  \left \| \mbs k  \right \| ^{-\alpha}
\widehat{\frac{ \dif B_t}{\sqrt{\Delta t}}}(\mbs k)
\text{ with } s = 3 - 2 \alpha =-\frac 5 3,
 \end{eqnarray}  
 where $A$ is a constant to ensure $
\Exp \left \| \bsigma_H \dot{ \B} \right \|^2 = 2 \frac {a_H}{\Delta t}$ (see equation \eqref{link energy dissipation in homogeneous case} above), $\widehat{\dif B}_t$ is the spatial Fourier transform of $\dif B_t$, with $\frac{ \dif B_t}{\sqrt{\Delta t}}$, a discrete scalar white noise process of unit variance in space and time. To sample the small-scale velocity, we first sample $\frac{ \dif B_t}{\sqrt{\Delta t}}$, to get $\widehat{\frac{ \dif B_t}{\sqrt{\Delta t}}}$, and finally $\widehat{\bsigma_H \dot{ \B}}(\mbs k)$ with the above equation.

\subsection{Resolution gain on a single simulation}
\label{Resolution gain on a single simulation}

\begin{figure}
\begin{center} 
    \textbf{Models comparison}\par\medskip
\includegraphics[width=15cm]{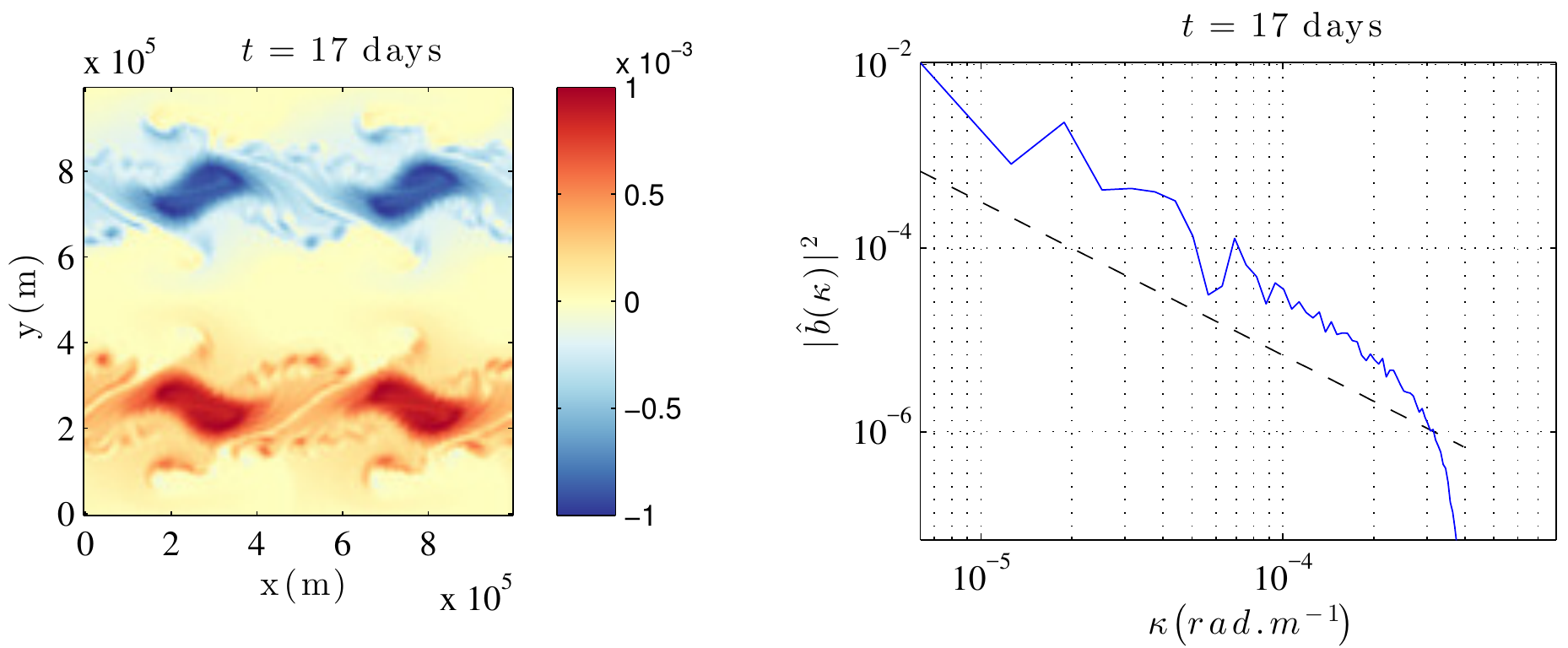}
\includegraphics[width=15cm]{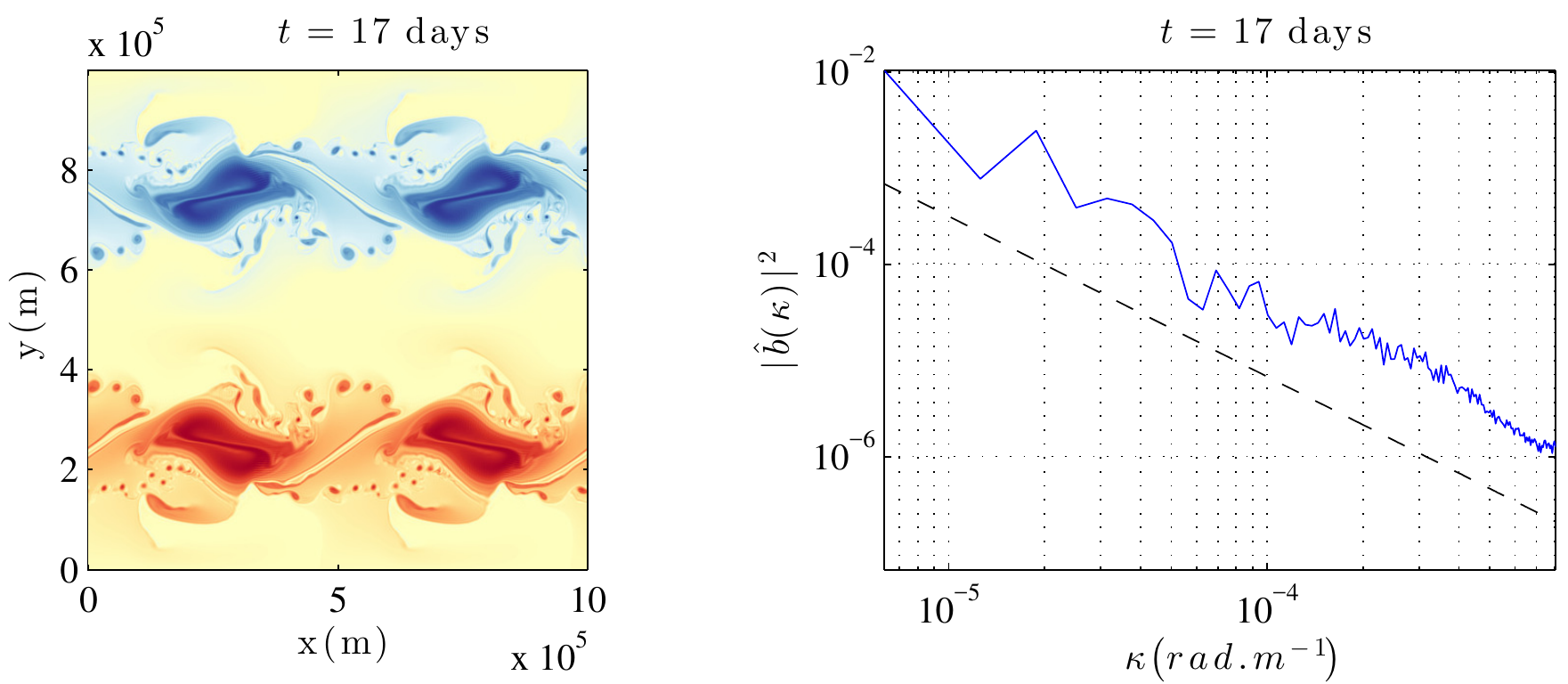}
\includegraphics[width=15cm]{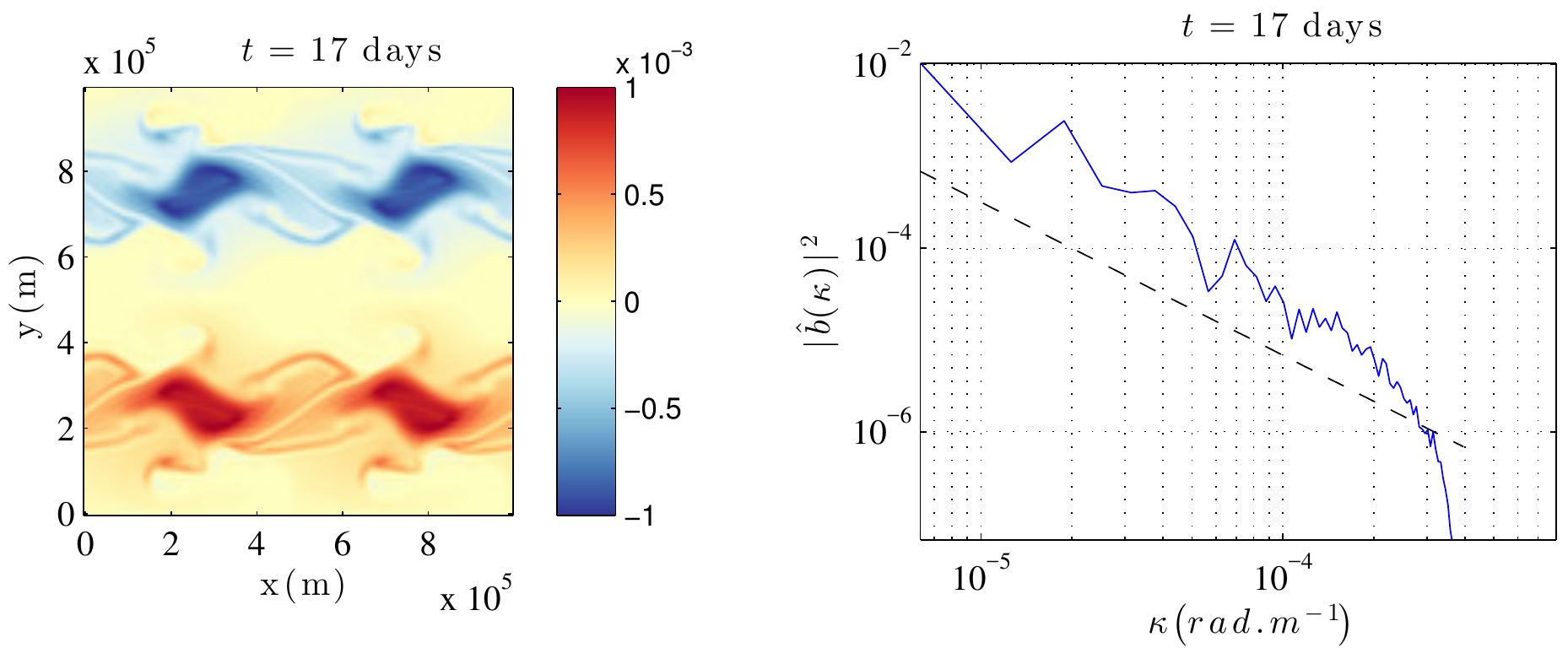}
\end{center}
\caption{Buoyancy ($m.s^{-2}$) and its spectrum ($m^2.s^{-4}/($rad$.m^{-1})$) at the $17^{th}$ day of advection for ${\rm SQG_{MU}}$ at resolution $128^2$ (top), SQG at resolution $512^2$ (middle) and  at resolution $128^2$ (bottom). Unlike ${\rm SQG_{MU}}$, the low-resolved  SQG simulation diffuses the ``pearl necklaces'', noticeable only at higher resolution.}
\label{compHR_pearl_neaklace}
\end{figure}

In Figure \ref{compHR_pearl_neaklace}, the buoyancy field and its spectrum for low resolution ${\rm SQG_{MU}}$ and deterministic SQG simulations are displayed for the day $17^{th}$. 
For the spectrum plots (right column), the slope $-\frac 5 3 $ is superimposed. While the spectrum tail of the SQG model falls slightly before the stochastic one, the most significant gain is observed in the spatial domain, i.e. in the phase of the tracer. Indeed, the ${\rm SQG_{MU}}$ buoyancy field exhibits pearl-necklaces, only obtained at higher resolution. The low-resolved SQG simulation only generates smooth and stable filaments. Though small-scale energy distribution remains similar for both low-resolved models, 
the phase of the stochastic tracer is more accurate. This may seem surprising since the unresolved velocity, $\bsigma_H \dot{\mbs B}$, is defined in a loose way, through its spectrum, without prescribing the nature of its phase. However, the noise is multiplicative, and the random forcing, $-(\bsigma_H \dot{\mbs B})\bcdot \nab b$, does implicitly take into account the tracer phase.

\begin{figure}
\begin{center} 
    \textbf{Prescribed noise variance}\par\medskip
\includegraphics[width=15cm]{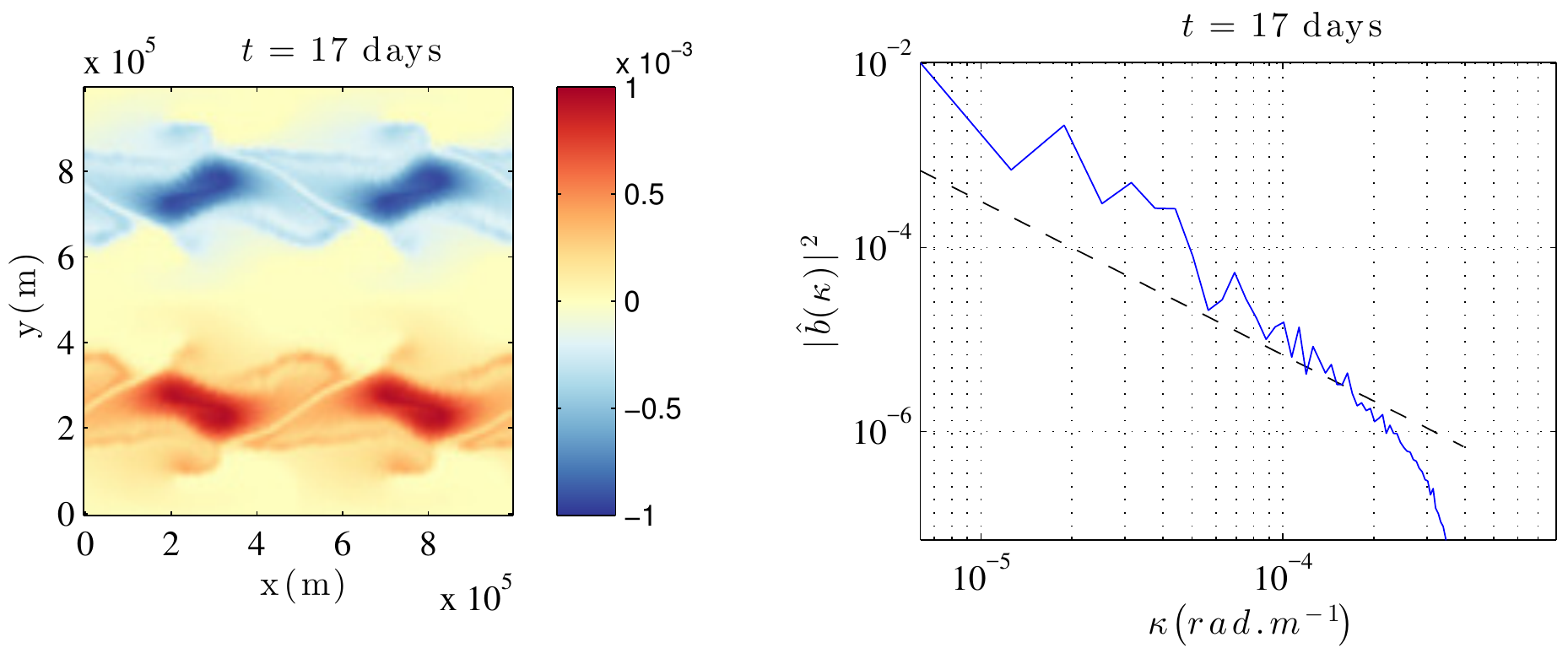}
\includegraphics[width=15cm]{images/sensitivity_noise/spectrum_SQGmodif_17_spot2_bruit_bon-eps-converted-to.pdf}
\includegraphics[width=15cm]{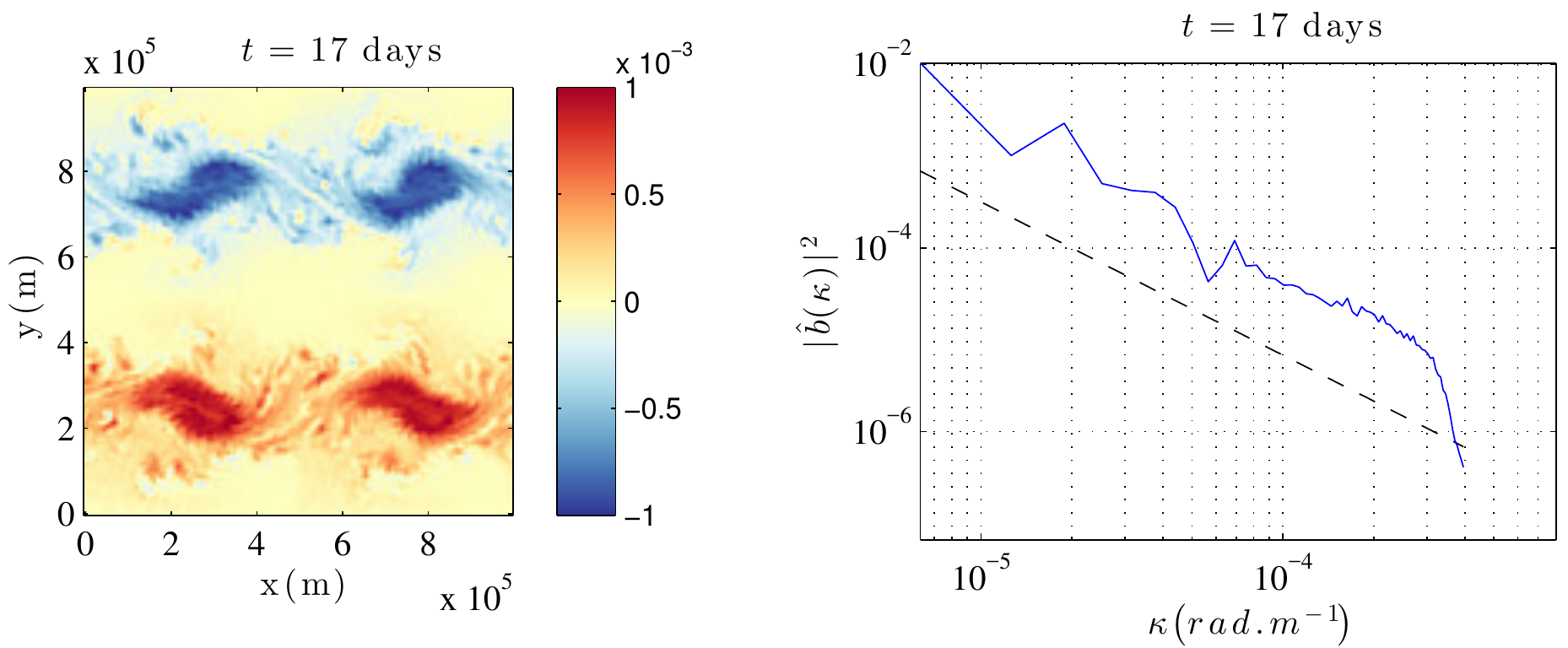}
\end{center}
\caption{Buoyancy ($m.s^{-2}$) and its spectrum ($m^2.s^{-4}/($rad$.m^{-1})$) at the $t=17^{th}$ day of advection for the ${\rm SQG_{MU}}$ model with a small-scale velocity component three times weaker than the one prescribed by the diffusion coefficient $a_H$ (top), with the correct amount of small-scale energy (middle) and a small-scale velocity three times higher than the model diffusion. If the prescribed balance between noise and diffusion is not met the tracer field becomes quickly too smooth or too noisy.}
\label{plot_sensitivity_noise}
\end{figure}

Note, within the stochastic framework, the diffusion coefficient is explicitly related to the noise variance. 
If the small-scale velocity is set to a magnitude three times smaller than the one prescribed by the diffusion coefficient $\frac{a_H} 2$, the tracer field becomes quickly too smooth (see Figure \ref{plot_sensitivity_noise}). Conversely, if the small-scale velocity is set to a magnitude three times larger than dictated by the stochastic transport model, the tracer field becomes rapidly too noisy. 
This is visible both in the spatial and Fourier spaces (Figure \ref{plot_sensitivity_noise}). The stochastic transport model thus imposes a correct balance between noise and diffusion.

\subsection{Ensemble forecasts}
\label{Improving for ensemble forecasts I: uncertainty quantification}

While single realization of ${\rm SQG_{MU}}$ model carries more valuable information than a deterministic SQG formulation at the same resolution, our model further 
enables to perform ensemble forecasting and filtering. Straightforwardly, an ensemble of independently randomly forced realizations $\{b^{(i)} | i=1,\cdots,N_e\}$ of tracer $b$ can be simulated according to the SPDE \eqref{transport SQG moderate uncertainty}. The probability density function and all the statistical moments of the simulated tracer can then be approximated. For instance, the (ensemble) mean of the buoyancy is a spatio-temporal field defined by:
\bea 
\Exp (b) (\xx,t) 
\approx 
\hat{\Exp} (b) (\xx,t) 
\defi
\frac 1 {N_{e}} \sum_{i=1}^{N_{e}}
b^{(i)}(\xx,t),
\eea 
where $N_{e}$ denotes the ensemble size. This is in essence a  Monte-Carlo Makov Chain (MCMC) simulation. The ensemble size is deliberately kept small\footnote{All the random simulations are performed with $200$ -- $128^2$ mesh-size -- realizations.} in order to assess the proposed stochastic framework skills.

We compare the ensemble bias with the estimated error provided by the ensemble itself. The bias corresponds to the discrepancy between the tracer ensemble mean and the SQG simulation at high resolution\footnote{Note this simulation is afterward spatially filtered and subsampled to the same resolution as the ensemble} ($512^2$).

Our reference is deterministic since the initial condition is perfectly known and the target dynamics is deterministic, as the real ocean dynamics. The partial knowledge of initial conditions is a complementary issue not addressed in this paper. The reference being deterministic, the bias represents both the error of the mean and the mean of the error:
\bea
 \widehat{\Exp}\{ b\} -  b^{ref} 
=
\widehat{\Exp}\{ \epsilon \}
,
\label{error mean}
\eea
where $\epsilon=
b -  b^{ref}
 $ stands for the (random) error. We denote by $e$ the absolute value of this bias. Another error metric could be the Root Mean Square Error (RMSE), $\sqrt{\widehat{\Exp}\{ \epsilon^2 \}}$. Yet slightly larger, it is found to have similar spatial and spectral distributions (not shown).

\begin{figure}
\begin{center} 
    \textbf{Spread-error consistency in the spatial domain}\par\medskip
\includegraphics[width=15cm]{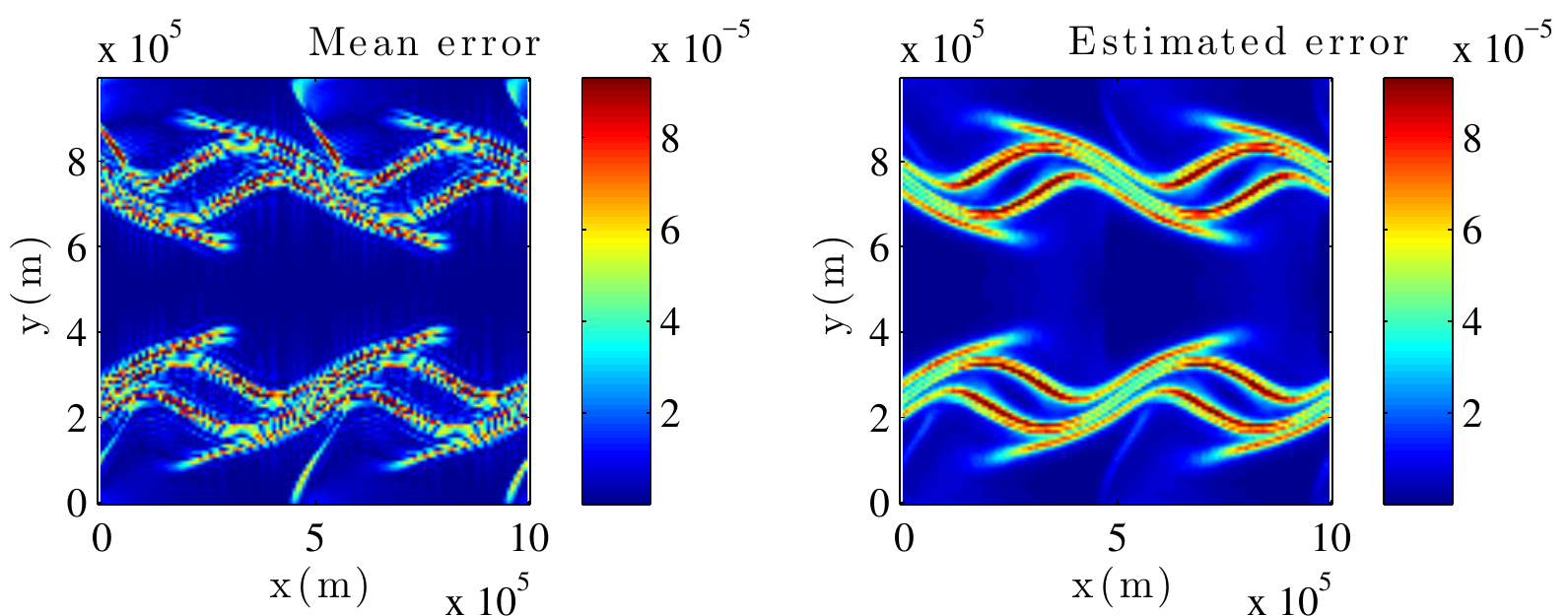}
\includegraphics[width=15cm]{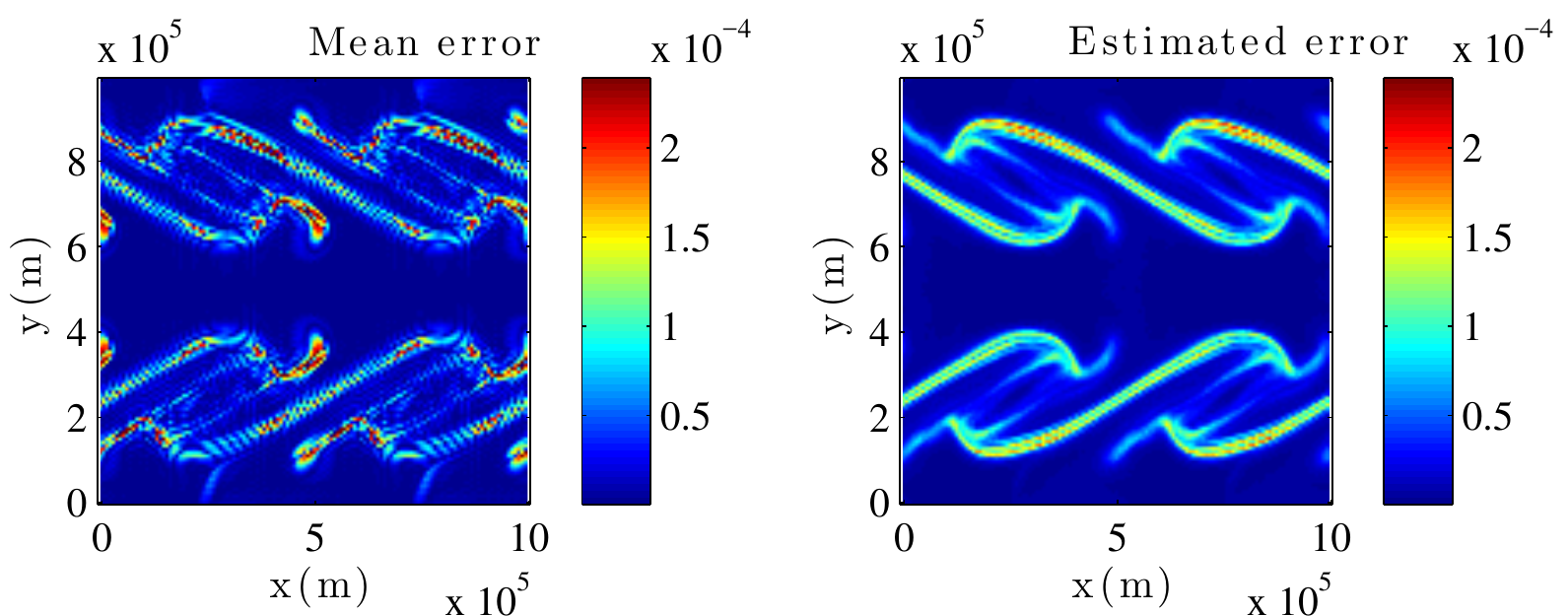}
\includegraphics[width=15cm]{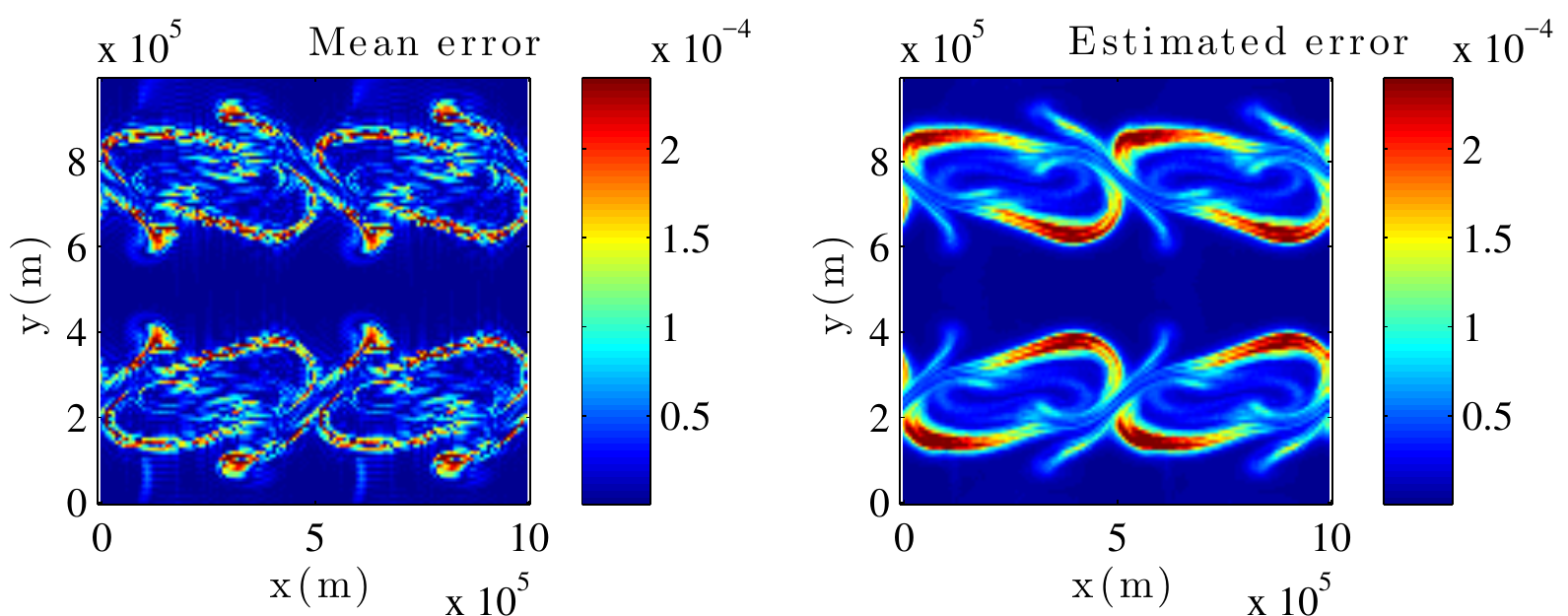}
\end{center}
\caption{Buoyancy bias absolute value, $e= |\widehat{\Exp}\{ b\} -  b^{ref} |$, ($m.s^{-2}$) of the ${\rm SQG_{MU}}$ model (left) and its estimation, $\epsilon_{est}$, ($1.96 \times$ the standard deviation of the ensemble) (right) at resolution $128^2$ at (from top to bottom) $t=10, 13$ and $15$ days of advection. The reference is the usual SQG model at resolution $512^2$ -- adequately filtered and subsampled.}
\label{estimated_spatial_error}
\end{figure}

\begin{figure}
\begin{center} 
    \textbf{Spread-error consistency in the spatial domain}\par\medskip
\includegraphics[width=15cm]{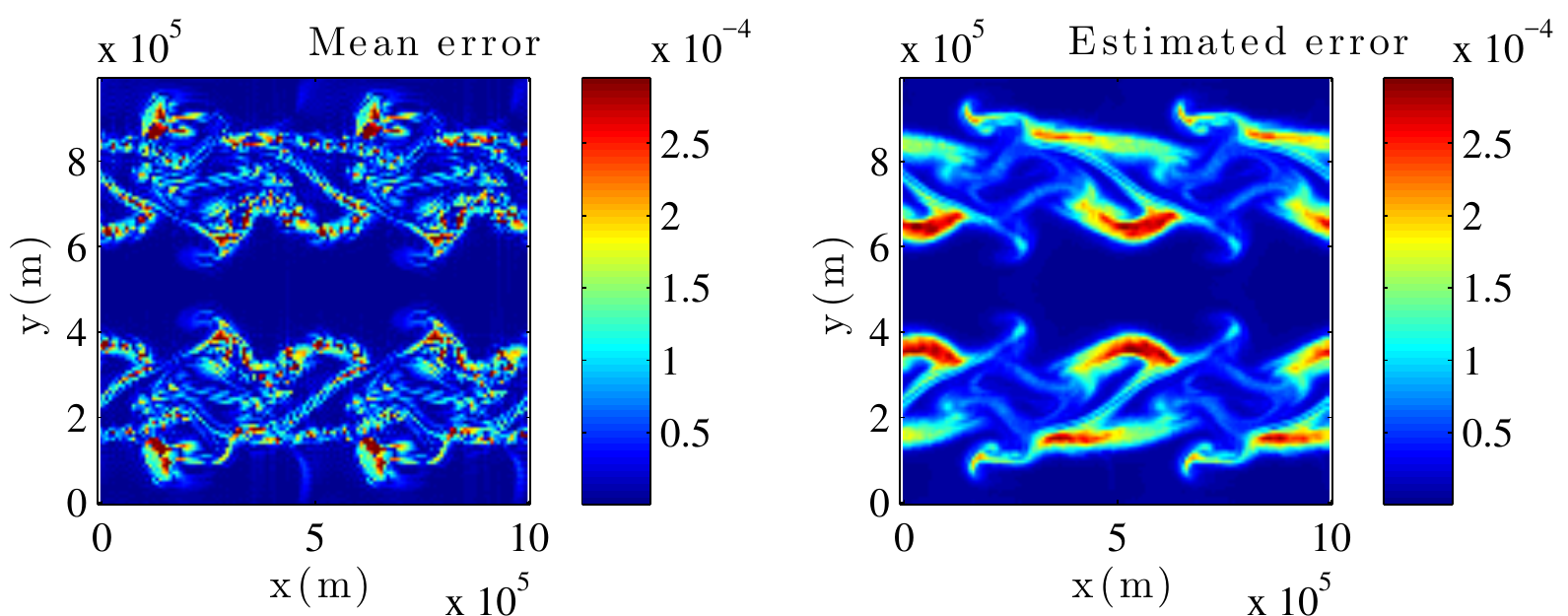}
\includegraphics[width=15cm]{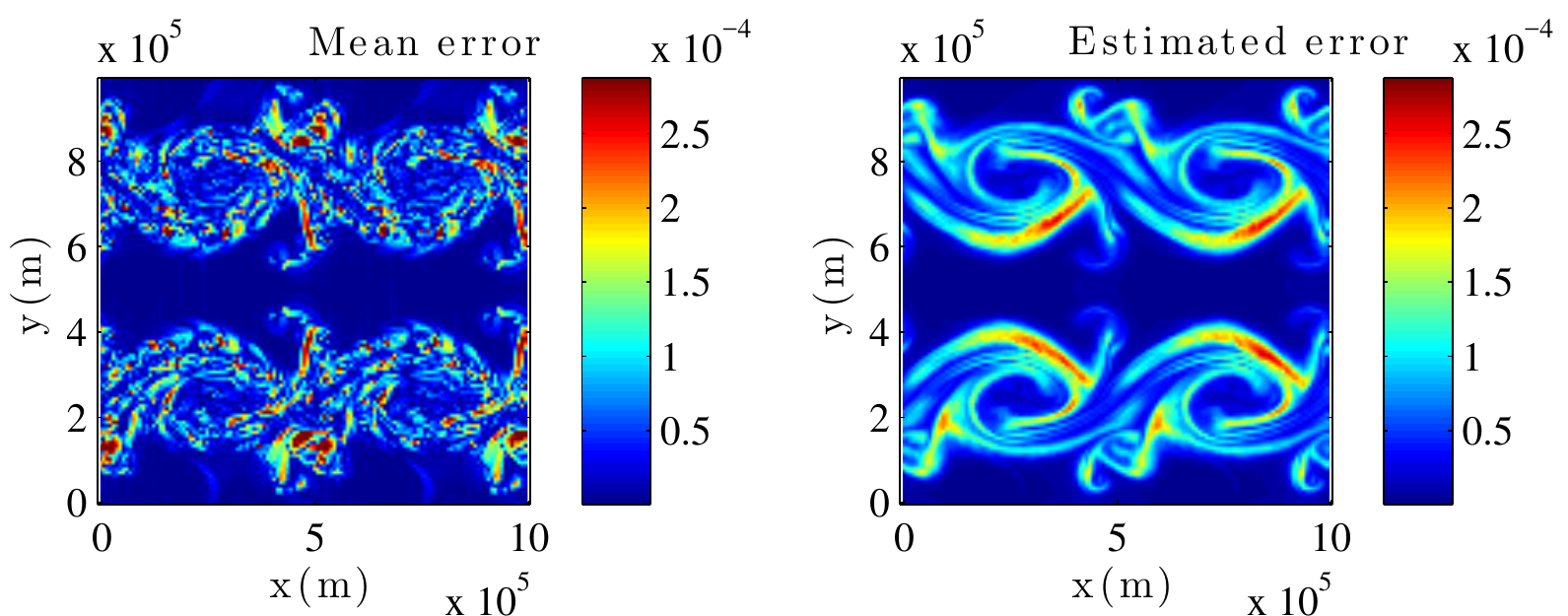}
\includegraphics[width=15cm]{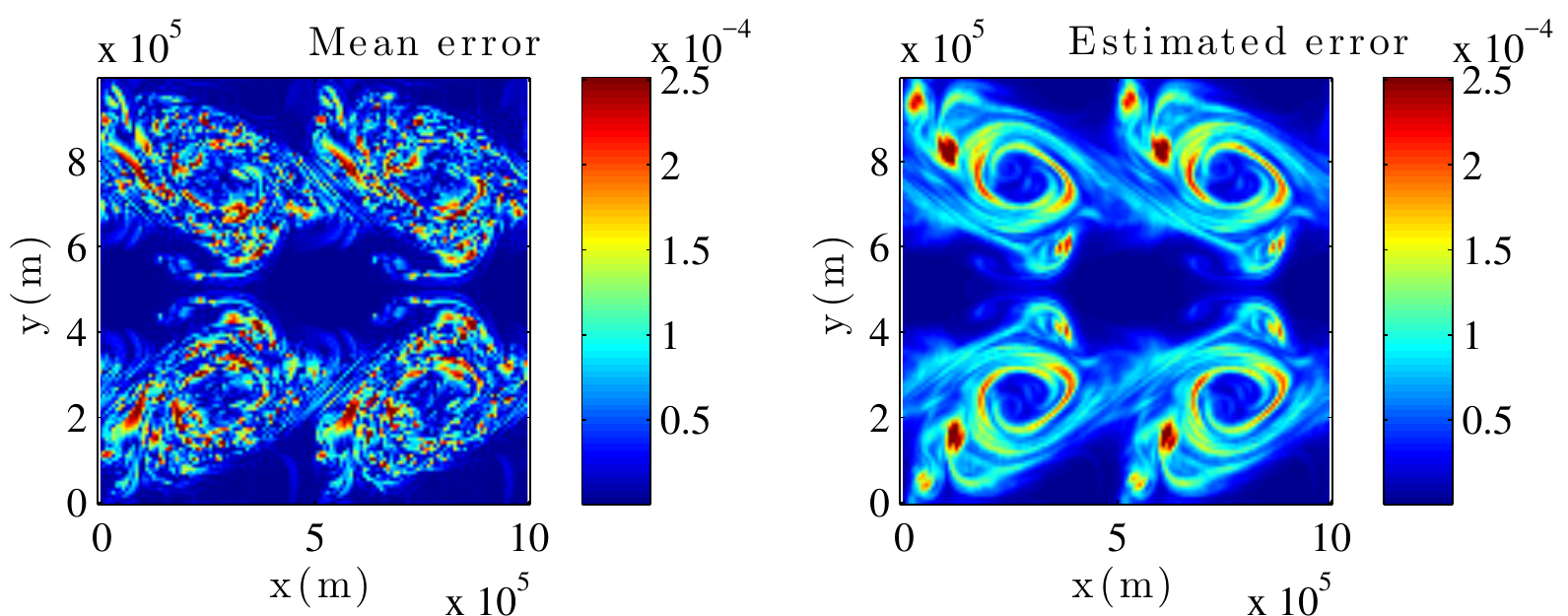}
\end{center}
\caption{Buoyancy  bias absolute value, $e= |\widehat{\Exp}\{ b\} -  b^{ref} |$, ($m.s^{-2}$) of the ${\rm SQG_{MU}}$ model (left) and its estimation, $\epsilon_{est}$, ($1.96 \times$ the standard deviation of the ensemble) (right) at resolution $128^2$ at (from top to bottom) $t=17, 20$ and $25$ days of advection. The reference is the usual SQG model at resolution $512^2$ -- adequately filtered and subsampled.}
\label{estimated_spatial_error2}
\end{figure}

The estimated error, denoted $\epsilon_{est}$, is set to $1.96$ times the ensemble standard deviation. This specific value corresponds to the (Gaussian) $95\%$  confidence interval. Although the tracer distribution is not Gaussian, this value provides an accurate conventional error estimate:
\bea 
\epsilon_{est}^2(\xx,t) 
= (1.96)^2 \widehat{Var}(b)
\defi (1.96)^2 \frac 1 {N_{e}-1} 
\sum_{i=1}^{N_{e}} \left ( b^{(i)}- \hat{\Exp} (b) \right )^2.
\label{error mean estim}
\eea
As this error depends on time and space, several comparisons are performed at several distinct times in both the spatial and Fourier domains. In Figures \ref{estimated_spatial_error} and \ref{estimated_spatial_error2}, the absolute value of spatial fields \eqref{error mean} and \eqref{error mean estim} ({\em i.e.} $e$ and $\epsilon_{est}$) are compared at days $10$, $13$, $15$, $17$, $20$ and $25$. As  obtained, the ${SQG_{MU}}$ model enables the ensemble to predict the positions and the amplitudes of its own errors with a very good accuracy. 

\begin{figure}
\begin{center} 
\includegraphics[width=15cm]{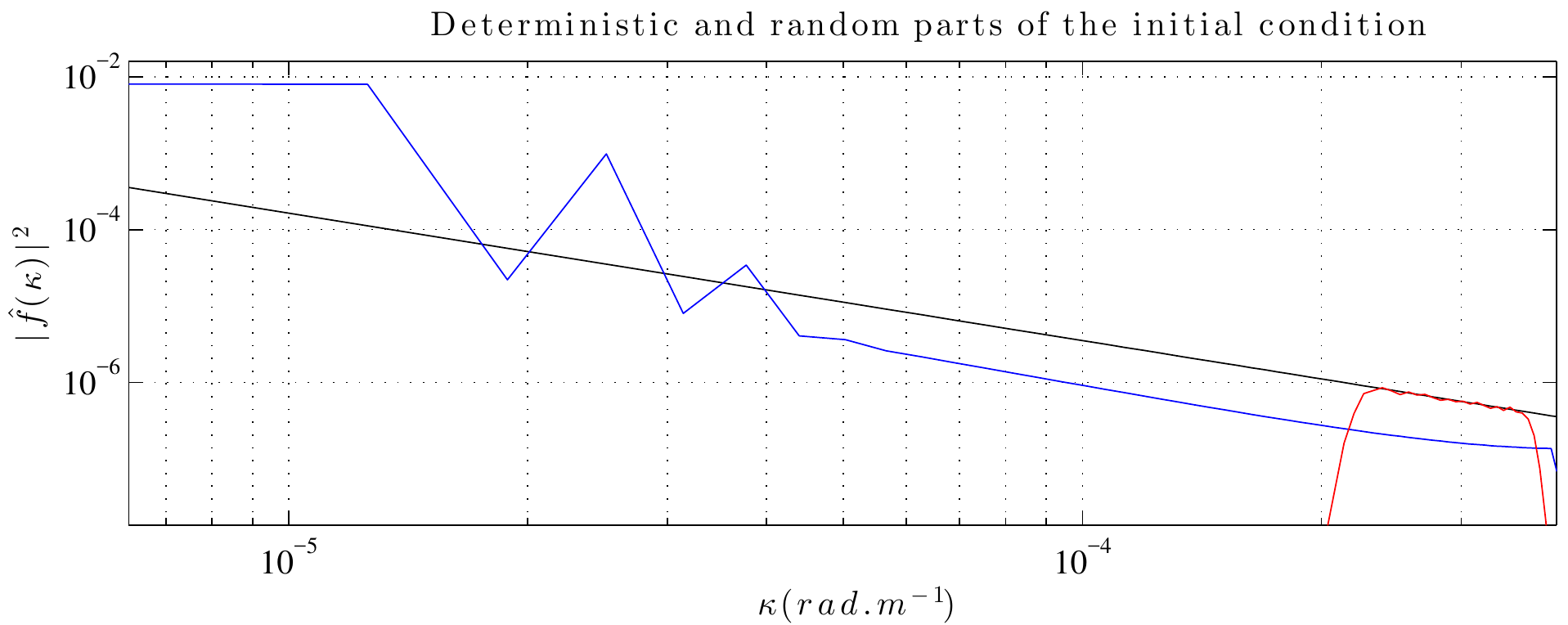}
\end{center}
\caption{Spectrum ($m^2.s^{-4}/($rad$.m^{-1})$), at the initial time, of the mean buoyancy, in blue, spectrum of its random perturbation, in red, and slope $- \frac 5 3$ in black. The initial perturbation is restricted to a narrow spectral band. This random initial condition has been used to simulate an ensemble with the deterministic SQG model.}
\label{spectrum_random_IC}
\end{figure}

To compare the spread-error consistency of the proposed model, a more classical type of random simulation is considered. An ensemble of the same size is initialized with random perturbations of the initial conditions \eqref{initial condition SQG simu}. The  perturbations are assumed to be homogeneous, isotropic, Gaussian and are sampled from a ($-\frac 53$) spectrum restricted to the small spatial scales, as shown in Figure \ref{spectrum_random_IC}. Then, the ensemble is forecast with the deterministic SQG model.

\begin{figure}
\begin{center} 
    \textbf{Spread-error consistency in the Fourier domain}\par\medskip
\includegraphics[width=15cm]{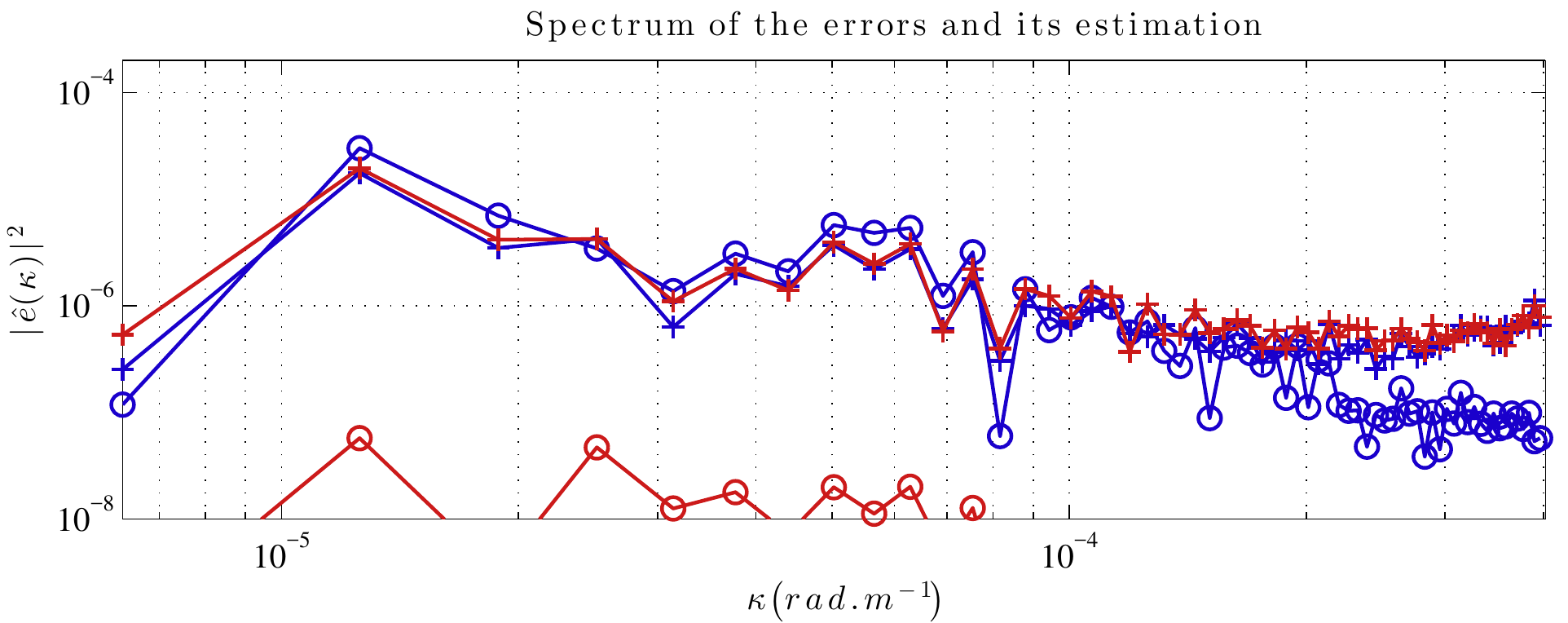}
\includegraphics[width=15cm]{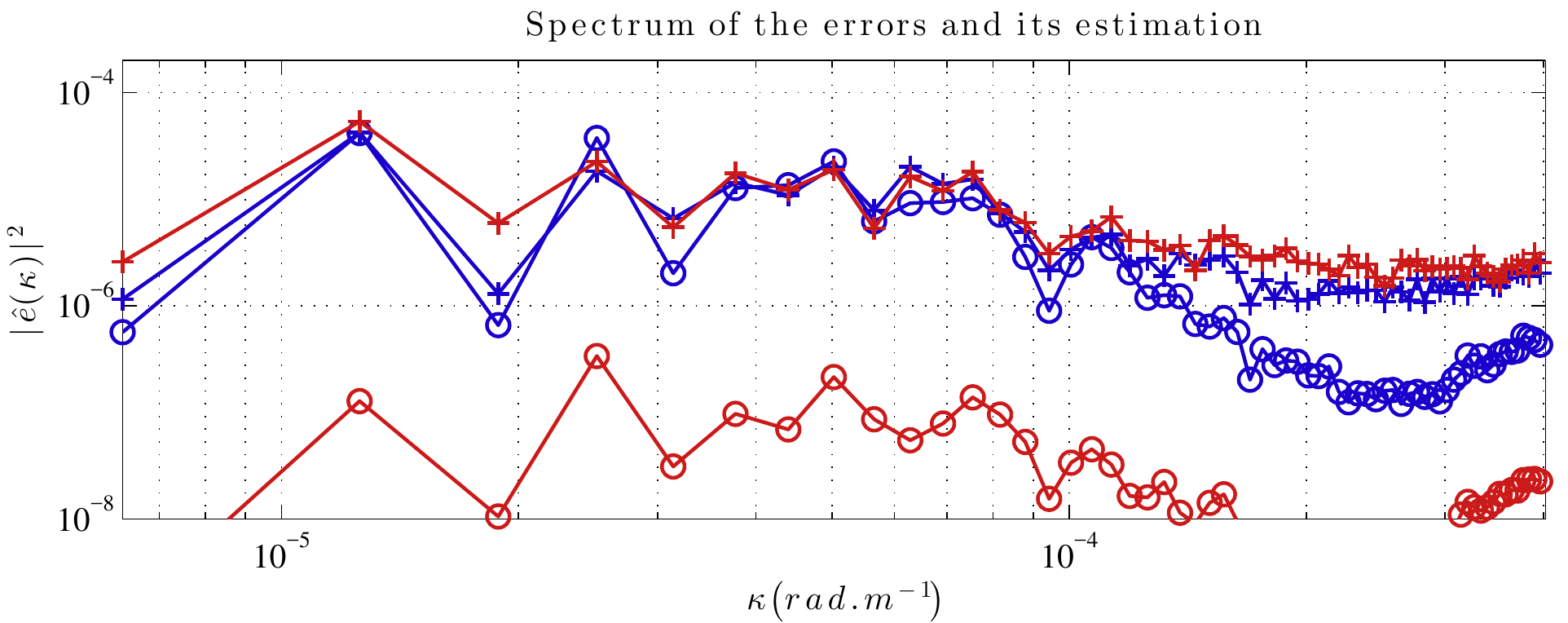}
\includegraphics[width=15cm]{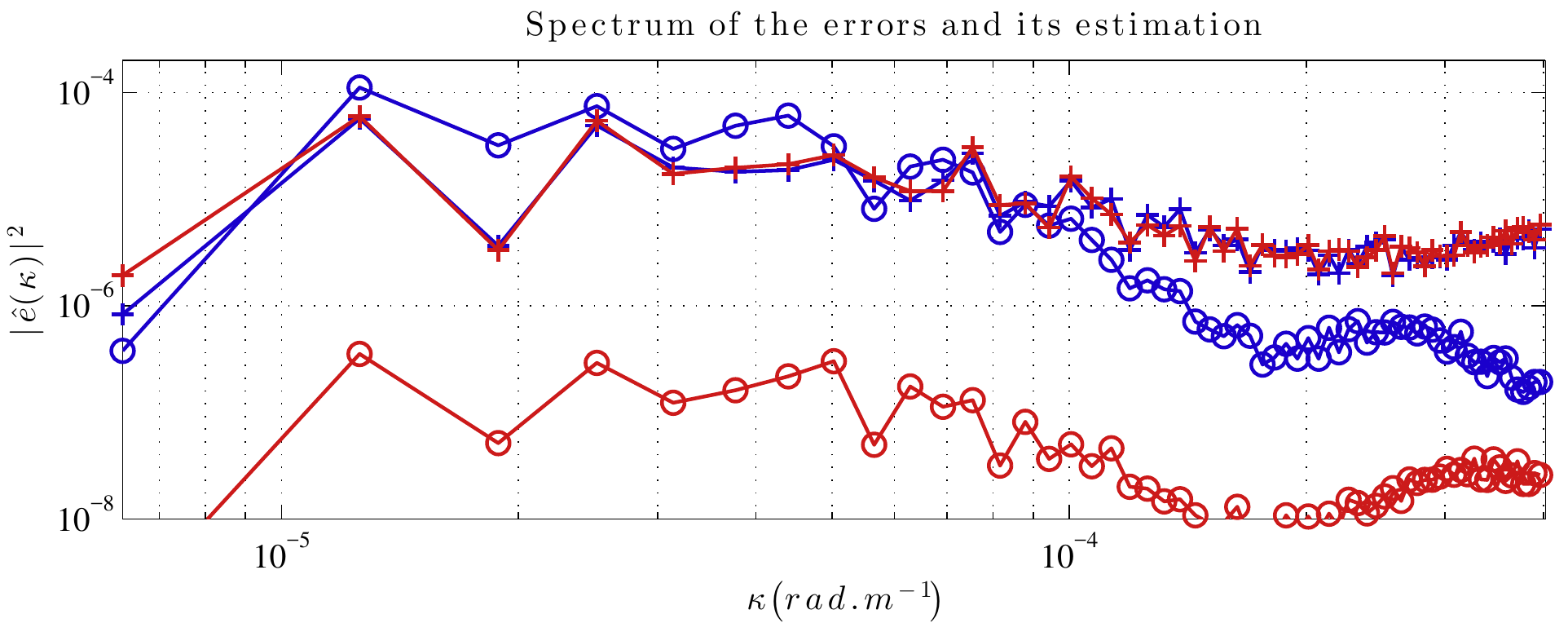}
\end{center}
\caption{Spectrum of the buoyancy bias absolute value, $e= |\widehat{\Exp}\{ b\} -  b^{ref} |$, (lines with crosses) and spectrum of the estimated error, $\epsilon_{est}$, ($1.96 \times$ the standard deviation of the ensemble) (lines with circles) ($m^2.s^{-4}/($rad$.m^{-1})$) of the low-resolution SQG model with random initial conditions (red) and of the ${\rm SQG_{MU}}$ model at the same resolution (blue), at (from top to bottom) $t=10, 13$ and $15$ days of advection. The reference is the usual SQG model at resolution $512^2$-- adequately filtered and subsampled. The low-resolution deterministic model with random initial conditions underestimates the error by at least one order of magnitude whereas our estimation is very precise except at small scales.}
\label{estimated_spectral_error}
\end{figure}

\begin{figure}
\begin{center} 
    \textbf{Spread-error consistency in the Fourier domain}\par\medskip
\includegraphics[width=15cm]{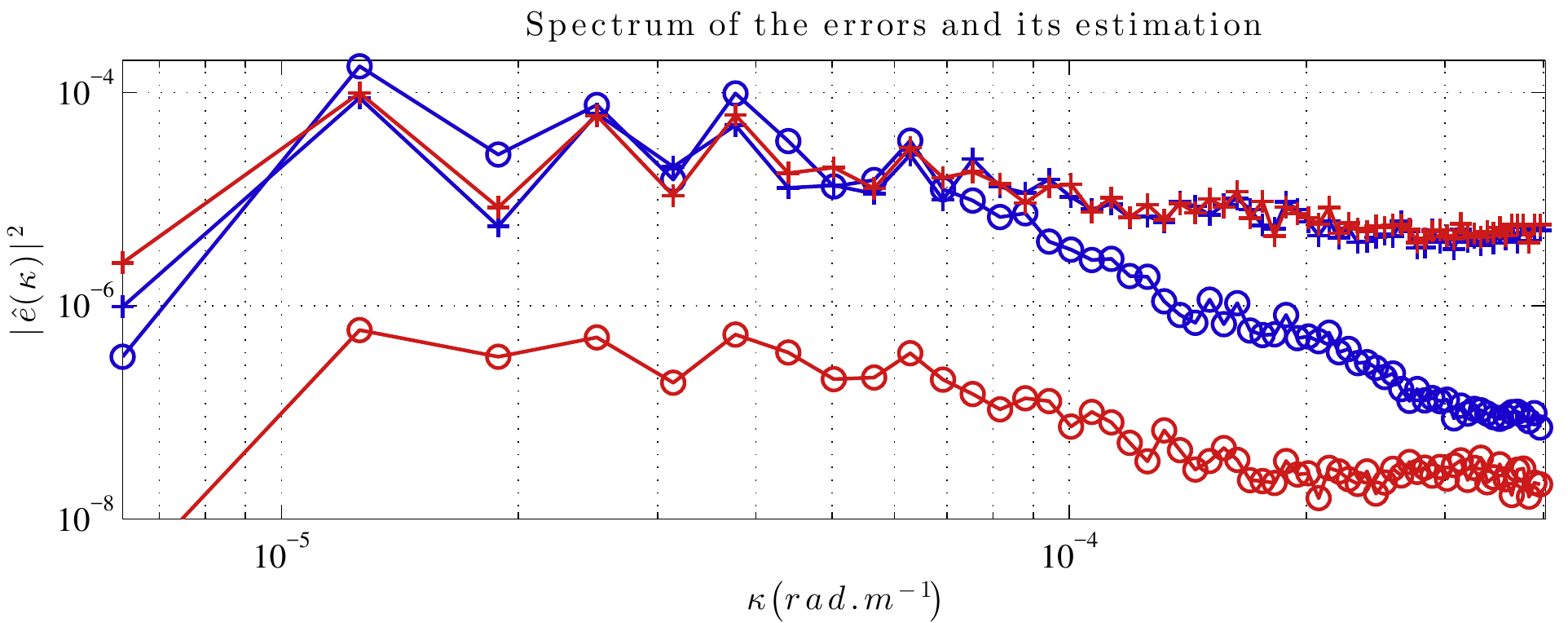}
\includegraphics[width=15cm]{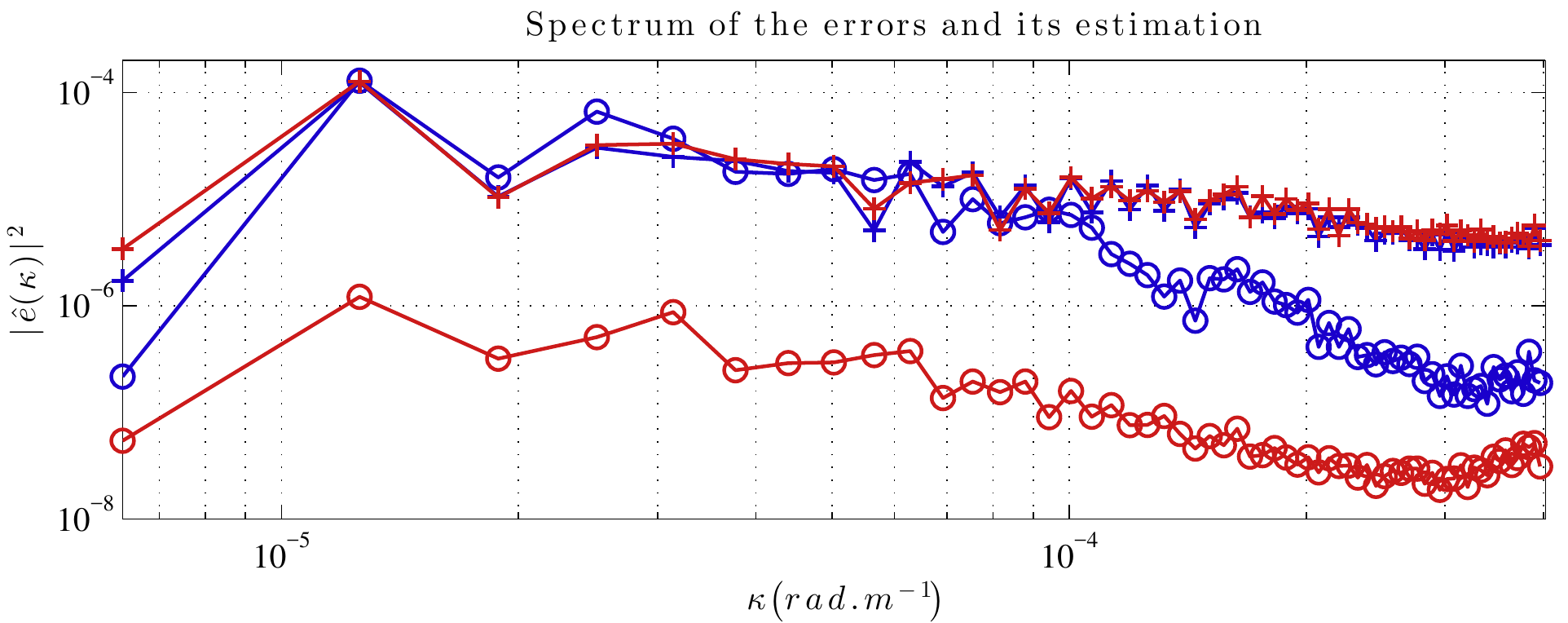}
\includegraphics[width=15cm]{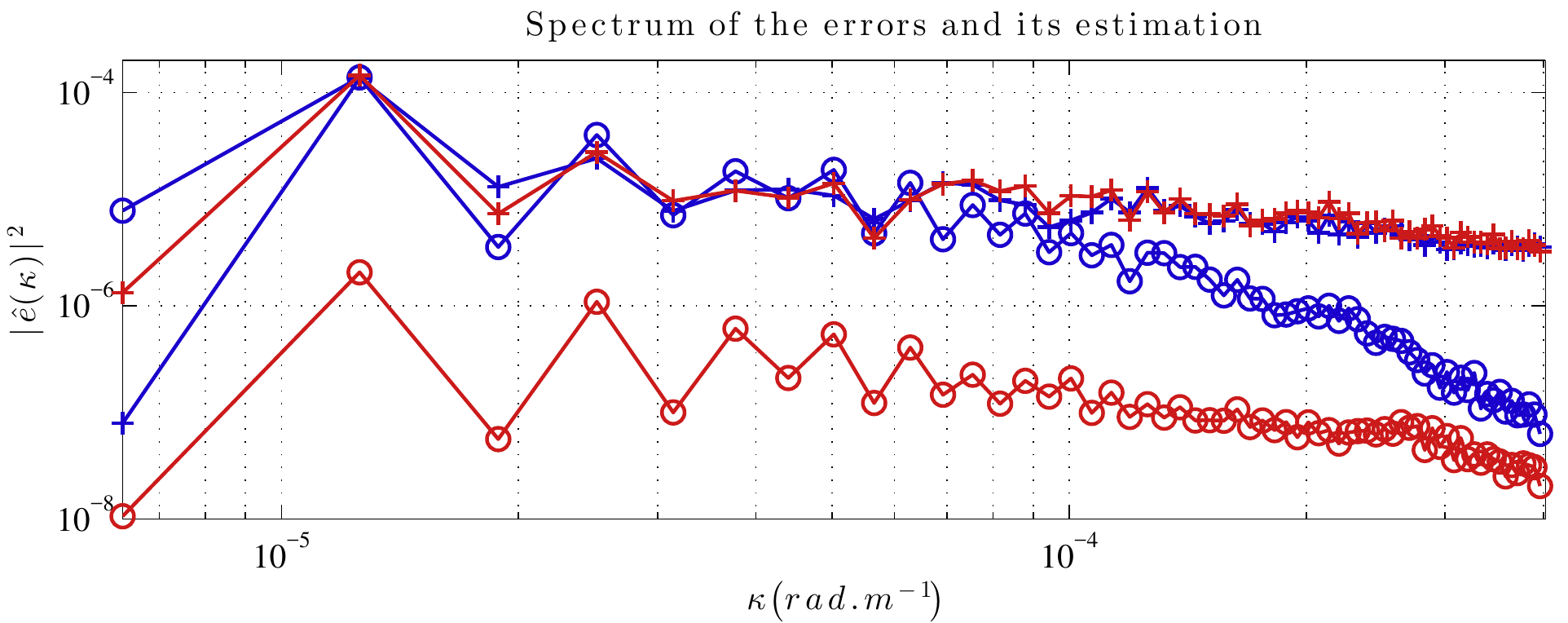}
\end{center}
\caption{Spectrum of the buoyancy bias absolute value, $e= |\widehat{\Exp}\{ b\} -  b^{ref} |$, (lines with crosses) and spectrum of the estimated error, $\epsilon_{est}$, ($1.96 \times$ the standard deviation of the ensemble) (lines with circles) ($m^2.s^{-4}/($rad$.m^{-1})$) of the low-resolution SQG model with random initial conditions (red) and of the ${\rm SQG_{MU}}$ model at the same resolution (blue), at (from top to bottom) $t=17, 20$ and $25$ days of advection. The reference is the usual SQG model at resolution $512^2$-- adequately filtered and subsampled. The low-resolution deterministic model with random initial conditions underestimates the error by at least one order of magnitude whereas our estimation is very precise except at small scales.}
\label{estimated_spectral_error2}
\end{figure}

Figures \ref{estimated_spectral_error} and \ref{estimated_spectral_error2} represent the spectrum of the errors. The blue and red lines with crosses stand for the spectrum of the 
bias absolute value, $e$, of the $SQG_{MU}$ with deterministic initial conditions and of the SQG model with random initial conditions, respectively. 

Deterministic and stochastic models have close distribution of errors over the scales, although the $SQG_{MU}$ ensemble mean generally leads to lower errors than the SQG ensemble mean. 

The blue line with circles denotes the spectrum of the $SQG_{MU}$ ensemble estimated error, $\epsilon_{est}$. As a benchmark, we superimposed the spectrum of the same estimator, $\epsilon_{est}$, but
simulated with the usual model (red curve with circles). 
This estimation is dramatically underestimated. It is generally one order of magnitude smaller that the real error. To reduce this drawback, a solution would be to multiply by $10$ the perturbations of the initial condition. However, this solution introduces strong errors on the realizations (not shown). Their small-scale errors are generally one order of magnitude larger than the ones of our model. These realizations of the deterministic model remain far from the reference for about ten days. On the contrary, the $SQG_{MU}$ predicts the correct spectral distribution of errors at each time, except at very small-scales, and each of its realizations are accurate as shown in the previous subsection.  Let us note however that most of the errors are concentrated at large scales.

$SQG_{MU}$ thus appears to provide a relevant ensemble of realizations, as it enables us to estimate the amplitude of its own error with a good accuracy both in the spatial and spectral domains.

\begin{figure}
\begin{center} 
    \textbf{$1^{st}$ and $2^{nd}$ point-wise moments}\par\medskip
\includegraphics[width=15cm]{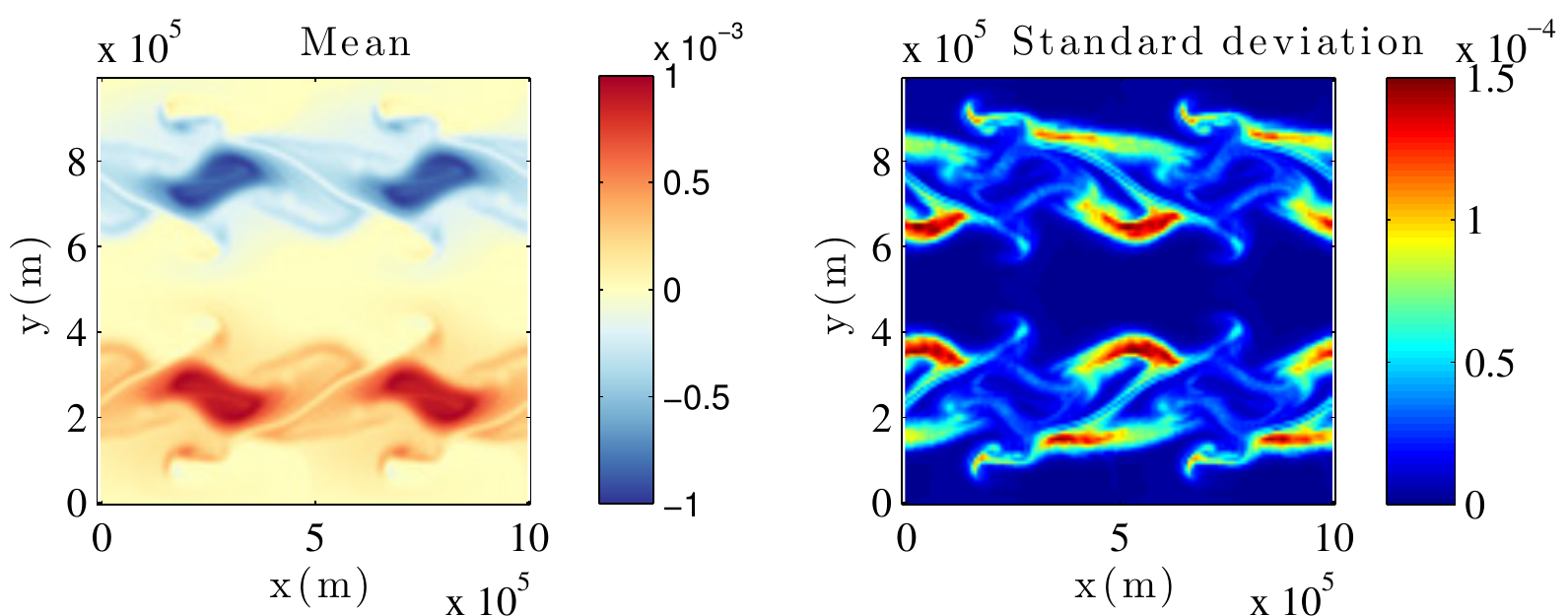}
\includegraphics[width=15cm]{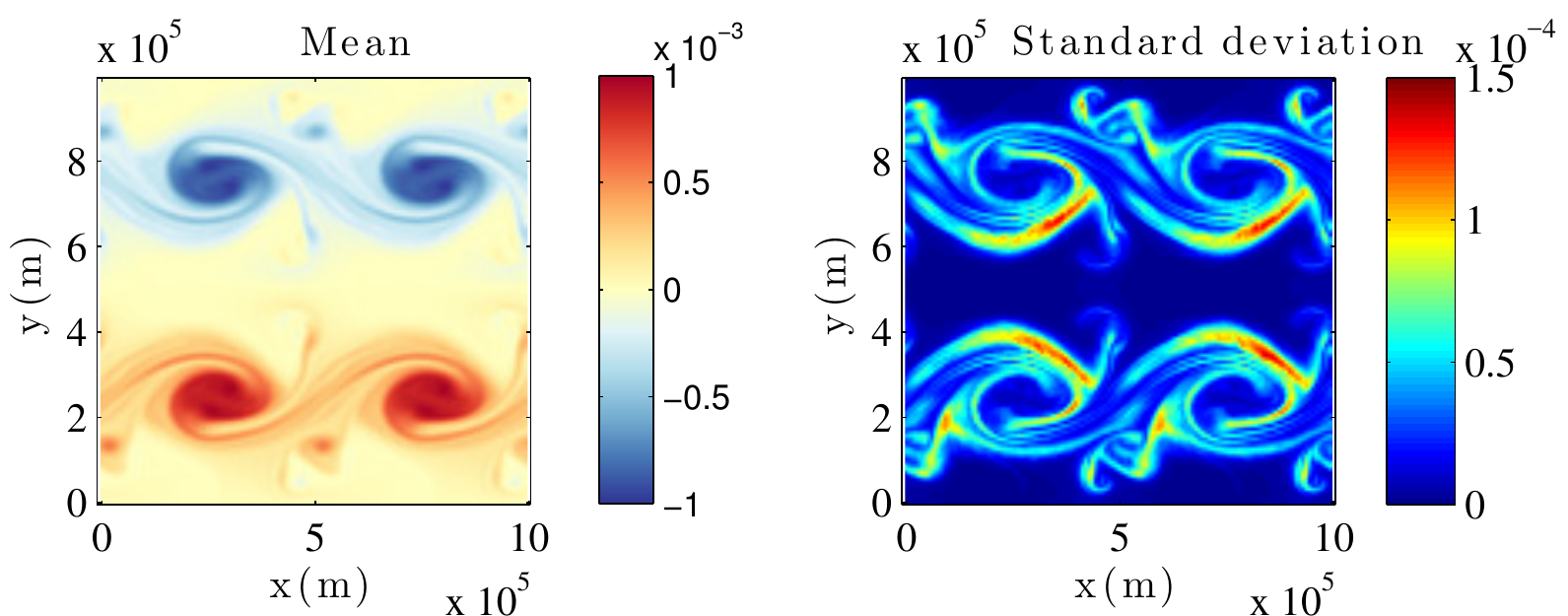}
\includegraphics[width=15cm]{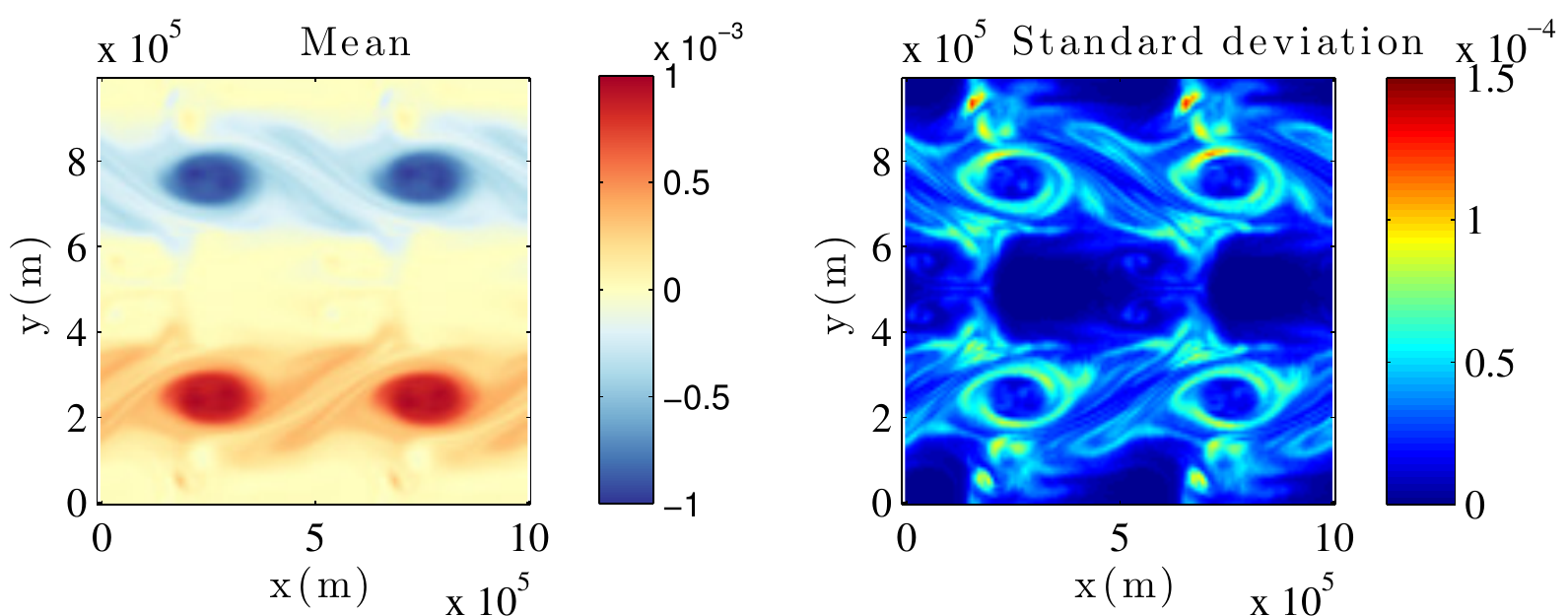}
\end{center}
\caption{Point-wise mean (left) and standard deviation (right) of the buoyancy ($m.s^{-2}$) at $t=17, 20$ and $30$ days of advection for $SQG_{MU}$ model at resolution $128^2$. The moments are computed through MCMC simulations. The mean field is smoother than the individual realizations. Areas of higher variance appear first near the stretched saddle points. Then, at $t=17$ days, the filament instabilities are triggered by the unresolved velocity component. The appearance of ``pearl necklaces" can be observed. At $t=20$, mushrooms-like structures also develop in the variance field near $(x,y)=(0,100),(500,100),(0,900)$ and $(500,900)$ (in km). At $t=30$ days, these random structures are transported by the zonal jets.}
\label{plot_spot2_moment12}
\end{figure}

With such an ensemble of realizations, it is now possible to  analyze the spatio-temporal evolution of the statistical moments.  In Figure \ref{plot_spot2_moment12}, we plotted  the ensemble tracer mean and variance for $t=17, 20$ and $30$ days of advection. As expected, the mean field is more smooth than the realizations (see Figure \ref{plot_sensitivity_noise} for comparison at $t=17$ days). One realization provides a more realistic field than the mean from a topological point of view. Indeed, the realization exhibits physically relevant small-scale structures. Nevertheless, those structures have uncertain shapes and positions. Therefore, on average, the mean field is closer (in the sense of the norm $\| \bullet \|^2_{L^2(\Omega)}$) to the reference. Besides, those uncertain small-scale structures, forgotten by the mean field, are visible in the variance. The variance becomes significant after $10$ days of advection, near the stretched saddle points. The strong tracer gradients create strong multiplicative noises. Indeed, strong large-scale gradients involve smaller scales, and thus interact with the small-scale velocity $\bsigma \dot{\mbs B}$. Then, at $t=17$ days, the filament instabilities are triggered by the unresolved velocity stretching effects. The appearance of ``pearl necklaces" and the underlying motions of those small-scale eddies are mainly determined by the action of the unresolved velocity component. In consequence, these structures are associated with a high uncertainty in their shapes and locations. Hence, they appear naturally on the variance field. At $t=20$, those sources of variance remain and mushroom-like structures also develop near $(x,y)=(0,100),(500,100),(0,900)$ and $(500,900)$ (in km). The evolution of these fronts are uncertain, and also show up in the variance field. On the day $30^{th}$, these random structures are transported by the zonal jets which are located at $y=0$ and $y=500$ km.

\begin{figure}
\begin{center} 
    \textbf{$3^{rd}$ and $4^{th}$ point-wise moments}\par\medskip
\includegraphics[width=15cm]{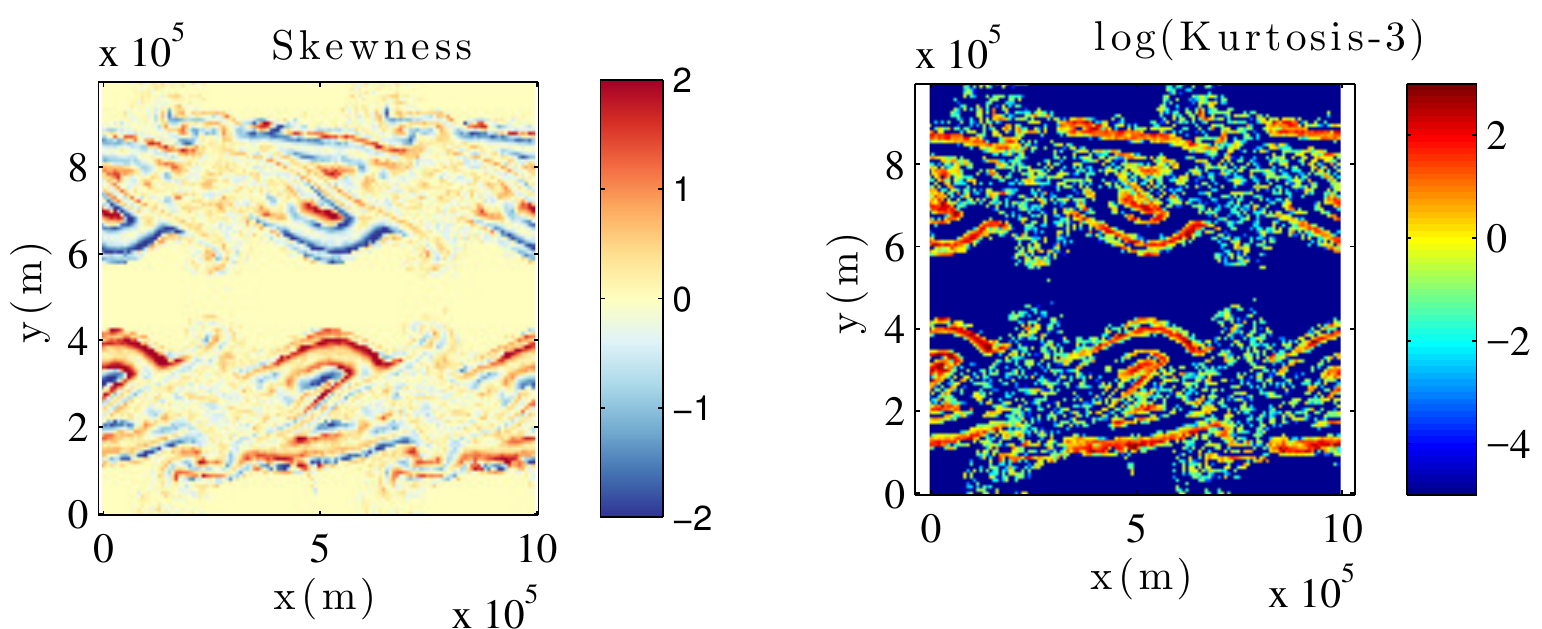}
\includegraphics[width=15cm]{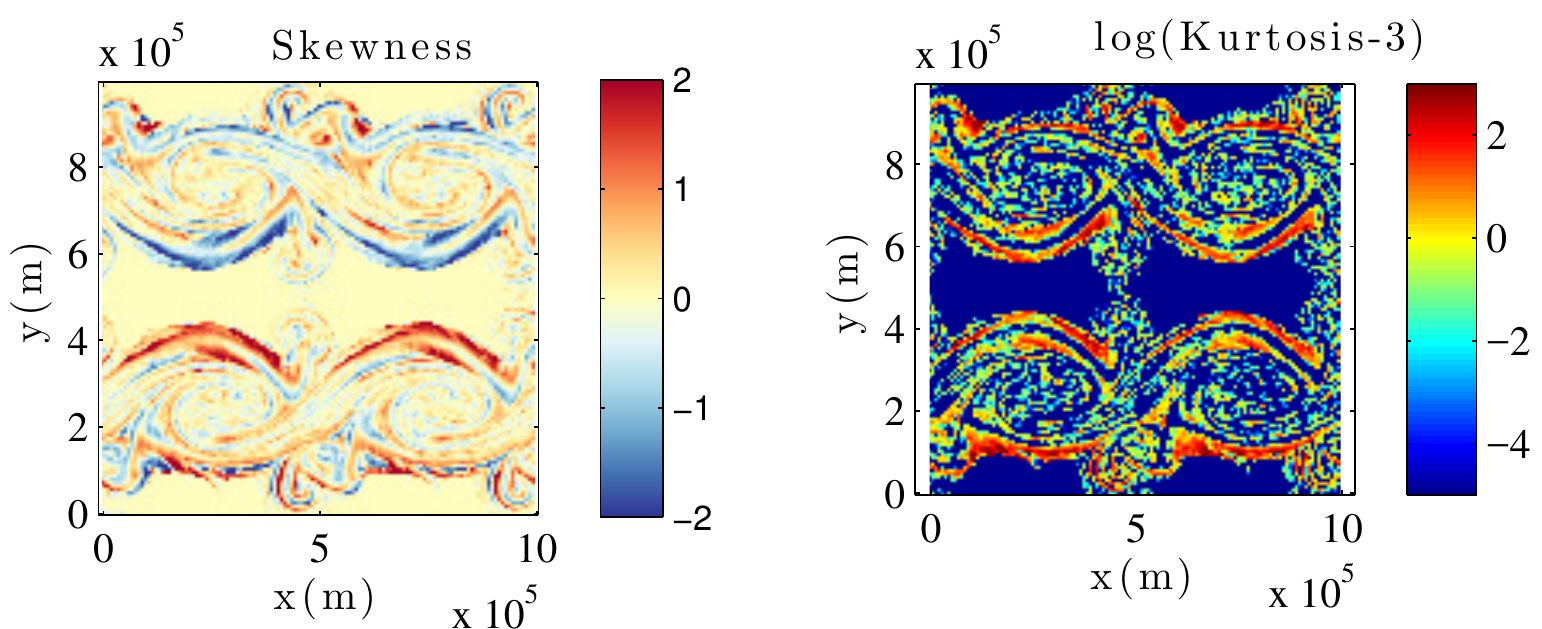}
\includegraphics[width=15cm]{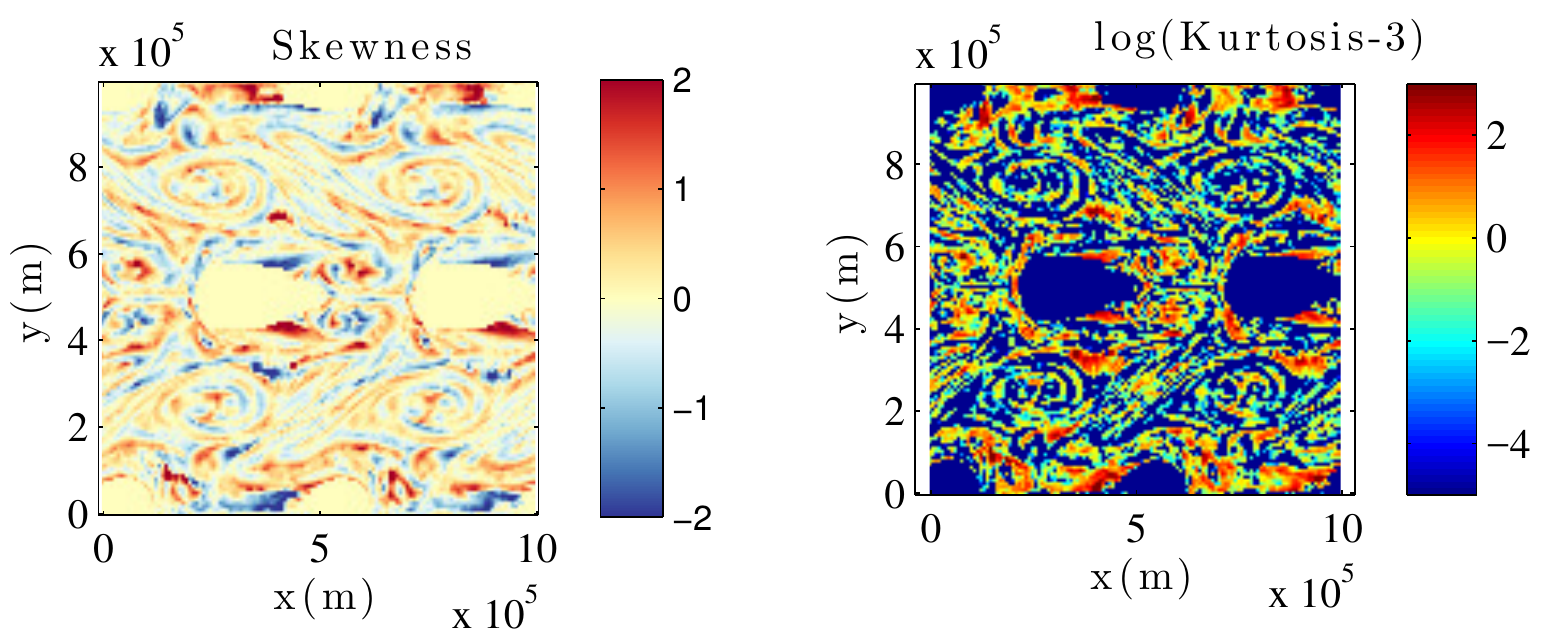}
\end{center}
\caption{Point-wise skewness, and logarithm of the excess kurtosis of the buoyancy at $t=17, 20$ and $30$ days of advection for ${\rm SQG_{MU}}$ model at resolution $128^2$. The moments are computed through MCMC simulations. Negative excess kurtosis is set to $0$. The point-wise law of the tracer is clearly non-Gaussian. The ``pearl necklace" events identified in the variance plots leads to  fat-tailed distributions with skewness at $t=17$ and $20$ days. The random structures, associated with fat tails are then trapped in the zonal jets.}
\label{plot_spot2_moment34}
\end{figure}

The empirical  moments of order $3$ and $4$ can also be evaluated with the ensemble. 
A high $4^{th}$ order moment 
directly relates to the occurrence of extreme events, which is very relevant for dynamical analysis. The point-wise $4$-th order moment is centered and normalized to obtain the so-called kurtosis:
\bea 
m_4 \defi
\frac {
\Exp \left( b - \Exp(b) \right)^4 
}{
\left( \Exp \left( b - \Exp(b) \right)^2  \right)^2 }.
\eea 
The excess kurtosis, $m_4-3$ highlights deviations from Gaussianity. In particular, positive values figure the existence of fat-tail distribution. On the right column of Figure \ref{plot_spot2_moment34}, the logarithm of the excess kurtosis is displayed for several distinct times. Negative values of the excess kurtosis (which indicates a flatter peak around the mean) have been set to zero. 
The ``pearl necklaces", identified in the variance plots, engender fat-tailed distribution at days $t=17$ and $20$. 
The small eddies of a ``pearl necklace" have similar vorticity and are close to each other, creating high shears between them. A given eddy can be ejected from the necklace by its closest neighbors, and led up to the north or south down. In such a case, the eddy reaches a zone of the space, neither warm nor cold, with weak variability (e.g. with both local mean and variance being low compared to eddy's temperature). This brings extreme tracer values in statistical homogeneous areas.  Finally, the random structures, associated with extreme events are trapped in the zonal jets.

The point-wise moment of order $3$ marks the asymmetry of the point-wise tracer distribution. The skewness is the third-order moment of the centered and normalized tracer:
\bea 
m_3 \defi 
\frac {
\Exp \left( b - \Exp(b) \right)^3 
}{
\left( \Exp \left( b - \Exp(b) \right)^2  \right)^{\frac 3 2} } .
\eea
Considering the interpretation of excess-kurtosis, the skewness identifies the predominant occurrence of cold (resp. warm) extreme events, associated with the cold (resp. warm) ``pearl-necklaces".

\section{Conclusion}

Models under location uncertainty involve a velocity partially time-uncorrelated. Accordingly, the material derivative, the interpretation of conservation laws, and the usual fluid dynamics models are modified. In this paper, the random Boussinesq model is approximated by the so-called QG equations.
 In our random framework, the approximation depends on 
sub-grid terms scaling. With moderate turbulent dissipation, the PV is randomly transported in the fluid interior up to three source/sink terms. Two of them are smooth in time and cancel out for homogeneous turbulence. The last forcing term -- a random enstrophy source -- is related to the angle between stable directions of resolved and unresolved velocities. Similarly to the deterministic case, a uniform PV yields a randomized SQG model, called $SQG_{MU}$, where the buoyancy is transported in the stochastic sense. 

Simulation results are considered for the $SQG_{MU}$ model which is a good representation of the transport under location uncertainty. As such, results are believed to hold for any fluid dynamics models under location uncertainty. As found, $SQG_{MU}$ better resolves small-scale tracer structures than a usual SQG model simulated at the same resolution. The prescribed balance between noise and diffusion has also been confirmed. 
As further highlighted, an ensemble of simulations was able to estimate the amplitude and the position of its own errors in both spatial and spectral domains. This result suggests that the proposed randomized dynamics should be well suited for filtering and other data assimilation methods. On the contrary, a deterministic model with randomized initial conditions, either creates strong errors in its realizations (one order of magnitude larger than the unperturbed deterministic dynamics), or underestimates its own errors (one order of magnitude too low). A MATLAB\textsuperscript{\sffamily\textregistered} code simulating the $SQG_{MU}$ model is available online (\url{http://vressegu.github.io/sqgmu}). 
\\

As a discussion, we can address the problem of uncertainty quantification (UQ) of an unresolved dynamics from an opposite point of view as the usual setting. Instead of specifying a form for the sub-grid velocity, we can wonder what is the optimal form of SPDE for UQ in fluid dynamics. As demonstrated, randomization of initial conditions is far from being sufficient to quantify uncertainty. 
Therefore, a random forcing is needed to inject randomness at each time step.
The simplest choice is a forcing uncorrelated in time. Otherwise, additional stochastic equations need to be simulated to sample a time-correlated process. This is not desirable in high dimension and the correlation time of the process is often small anyway \citep{berner2011model}. A forcing uncorrelated in time is a source of energy. So, to be physically acceptable, the SPDE should involve a dissipative term to exactly compensate this source, even in non-stationary regime. The simplest choices of dissipation are diffusion and linear drag. For small-scale processes, the first is more suitable. Now, what is the form of a noise which brings as much energy as a diffusion removes? The proposed approach constitutes a suitable solution toward this goal.
\\

To further improve the accuracy of the UQ, spatial inhomogeneity of the variance tensor $\mbs a$ can be introduced from data or from additional models, as discussed in \cite{resseguier2016geo1}. This inhomogeneity may reduce possible spurious oscillations of tracer stable isolines. Such oscillations are visible on Figure \ref{compHR_pearl_neaklace} on the sides of the largest vortices. The assumption of time decorrelation may also be a limitation. Nevertheless, 
as shown by the numerical simulations,
the method already achieve very good outcomes with an homogeneous noise component and no real time-scale separation between the resolved and unresolved velocities.
Note in particular that since the noise is multiplicative, the random forcing is inhomogeneous even for homogeneous small-scale velocity.

\cite{resseguier2016geo3} focuses on a system with a clear time-scale separation between the meso and sub-meso scale dynamics to explore the consequences of the QG assumptions under a strong uncertainty assumption ($\Upsilon \sim \Ros$). A zero PV directly appears in the fluid interior and the horizontal velocity becomes divergent. This divergence provides a simple diagnosis of the frontolysis on warm sides of fronts and frontogenesis on cold sides of fronts.

Future works shall also focus on the potential benefits of the stochastic transport for data assimilation issues. As foreseen, the proposed stochastic formalism opens new horizons for ensemble  forecasting techniques and other UQ based dynamical approaches \citep[e.g.][]{ubelmann2015dynamic}. This stochastic setup has also been used to characterize chaotic transitions associated with breaking symmetries, also demonstrating interesting perspectives in that context.  



\section*{Acknowledgments}

The authors thank Aur\'elien Ponte, Jeroen Molemaker and Jonathan Gula for helpful discussions. We also acknowledge the support of the ESA DUE GlobCurrent project, the ``Laboratoires d'Excellence''  CominLabs, Lebesgue and Mer through the SEACS project.


\bibliographystyle{plainnat}

\bibliography{biblio}


\appendix

\section{Non-dimensional Boussinesq equations}
\label{appendix Nondimensionalized Boussinesq equations}

To derive a non-dimensional version of the Boussinesq equations under location uncertainty \citep{resseguier2016geo1}, we scale the horizontal coordinates $\tilde \xx_h= L \xx_h$, the vertical coordinate $\tilde z=h z$, the aspect ratio $D=h/L$ between the vertical and horizontal length scales. A characteristic time $\tilde t=Tt$ corresponds to the horizontal advection time $U/L$ with horizontal velocity $\tilde{\mbs  u} =U\mbs u$. A vertical velocity $\tilde w=(h/L)U w$ is deduced from the divergence-free condition. We further take a scaled buoyancy $\tilde b = B b$, pressure $\tilde \phi'=\Phi \phi'$ (with the density scaled pressures $\phi'=p' / \rho_b$ and $\dif_t \phi_{\sigma}=\dif_t  p_{\sigma}/\rho_b$), and the earth rotation $\mbs f^*= f\mbs k$. For the uncertainty variables, we consider a horizontal uncertainty $\tilde {\mbs  a}_H =  A_u\;\mbs a_H$ corresponding to the horizontal $2\times 2$ variance tensor; a vertical uncertainty vector $\tilde{a}_{z z}= A_w { a}_{z z}$ and a horizontal-vertical uncertainty vector $\tilde{\mbs a}_{H z}= \sqrt{A_u A_w} {\mbs a}_{H z}$ related to the variance between the vertical and horizontal velocity components. The resulting non-dimensional Boussinesq system under location uncertainty becomes:

\fcolorbox{black}{lightgray}{
\begin{minipage}{0.95\textwidth}
\begin{center}
\bf  Nondimensional  Boussinesq equations under location uncertainty
\end{center}
\begin{subequations}
\label{sto-Boussinesq-buo}
\begin{align}
&\!\!\text{\em Momentum equations} \nonumber\\
\label{sto-Boussinesq-buo:Momentum equations}
&\;\;\;\;
\dif_t\mbs u 
+ \left(\w \bcdot \nab \right) \mbs u \dif t
+  \frac{1}{\Upsilon^{1/2} }\left( \bsigma_H \dBt  \bcdot \nab_H \right) \mbs u
+  \left( \frac \Ros {\Bu\Upsilon^{1/2}} \right)  (\bsigma \dBt)_z   \partial_z \mbs u
\nonumber \\ & \hspace{0.8cm}
-  \frac{1}{2 \Upsilon} \sum_{i,j\in H}  \partial_{ij}^2\biggl (a_{ij} \mbs u\biggr)  \dif t
+  O \left( \frac \Ros {\Upsilon \Bu} \right) 
+\frac{1}{\Ros}\left( f_0 + \Ros \beta y \right)  \mbs k  \times 
\left( \mbs u \dif t +  \frac{1}{\Upsilon^{1/2} }\bsigma_H \dBt \right)
\nonumber \\ & \hspace{0.8cm}
=
- \Eu\;\nab_H \left( \phi' \dif t + \frac{1}{\Upsilon^{1/2} } \dif_t \phi_\sigma \right) 
,\\
&\;\;\;\;
\dif_t w
+ \left(\w \bcdot \nab \right) w \dif t
+  \frac{1}{\Upsilon^{1/2} }\left( \bsigma_H \dBt  \bcdot \nab_H \right) w
+  \left( \frac \Ros {\Bu\Upsilon^{1/2}} \right)  (\bsigma \dBt)_z   \partial_z w
\nonumber
 \\ & \hspace{0.8cm}
-  \frac{1}{2 \Upsilon} \sum_{i,j\in H}  \partial_{ij}^2\biggl (a_{ij} w \biggr)  \dif t
+  O \left( \frac \Ros {\Upsilon \Bu} \right) 
=
\frac{\Gamma}{ D^2} b\dif t
  - \frac \Eu {D^2}\;\partial_z\left( \phi' \dif t + \frac{1}{\Upsilon^{1/2} } \dif_t \phi_\sigma \right) , 
\label{w-compad1}\\
&\!\!\text{\em Buoyancy equation} \nonumber\\
&\;\;\;\;
\nonumber
{\dif_t b} 
+  \left( \w_\Upsilon^*\dif t +   \frac{1}{\Upsilon^{1/2} }(\bsigma d\B_{t})\right)  
\bcdot \nab b
-\frac{1}{2}\frac{1}{\Upsilon} \nab_H \bcdot \bigl ( \mbs a_H \nab b \bigr ) \dif t
+ O \left( \frac \Ros {\Upsilon \Bu} \right) 
\\
& 
\hspace{4.5cm} +  \frac{1}{(\Fr)^2}\frac{1}{\Gamma} 
\left (
w^*_{\Upsilon/2} \dif t 
+ \left( \frac \Ros \Bu \right) \frac{1}{\Upsilon^{1/2}} (\bsigma \dif \B_t)_z
\right ) =0,\\
&\!\!\text{\em Effective drift} \nonumber\\
&\;\;\;\;
\w_{\Upsilon}^*
= 
\bigl(\mbs u^*_\Upsilon, w^*_\Upsilon \bigr)\transp ,
\nonumber\\
&\;\;\;\;\;\;\;\;
=
\left( 
\left( 
\w -\frac{1}{2\Upsilon} \nab\bcdot \mbs a_H
\right ),
\left( 
w 
- \left( \frac \Ros {2\Upsilon \Bu} \right)   \nab_H \bcdot {\mbs a}_{H z}
+ O \left( \frac \Ros {\Upsilon \Bu} \right)^2 
\right )
\right ) \transp,
\\
&\!\!\text{\em Incompressibility} \nonumber\\
&\;\;\;\;  \nab \bcdot \w =0,\\
&\;\;\;\; 
{\nab} {\bcdot}\bigl(\bsigma \dif \B_t\bigr)  = 0 ,\\
&\;\;\;\;  
 \nab_H \bcdot \left( \nab_H \bcdot \mbs a_H \right) \transp
 + 2 \frac \Ros \Bu \nab_H \bcdot \partial_z \mbs a_{Hz}
 + O \left( \left( \frac \Ros \Bu \right)^2 \right)
 =0.
\end{align}
\end{subequations}
\end{minipage}
} 
 \;\\\;\\

Here, we do not separate the time-correlated components and the time-uncorrelated components in the momentum equations.
The terms in $O\left( \frac \Ros \Bu \right)$ and $O\left( \frac \Ros \Bu \right)^2$ are related to the time-uncorrelated vertical velocity. These terms are too small to appear in the final QG model ($\Bu = O\left( 1\right)$ in QG approximation) and not explicitly shown. We only make appear the big $O$ approximations. Traditional non-dimensional numbers are introduced : the Rossby number $R_o= U/(f_0 L)$ with $f_0$ the average Coriolis frequency;
 the Froude number ($\Fr= U/(Nh)$), ratio between the advective time to the buoyancy time; $\Eu$, the Euler number,  ratio between the pressure force and the inertial forces, $\Gamma=Bh/U^2=D^2 B T/W$ the ratio between the mean potential energy to the mean kinetic energy. To scale the buoyancy equation, the ratio between the buoyancy advection and the stratification term has also been introduced:
 \begin{equation}
 \frac{B/T}{N^2W}
 =
 \frac{B}{N^2h}
 =
 \frac{U^2}{N^2h^2}
 \frac{Bh}{U^2}
 =
 Fr^2 \Gamma. 
 \end{equation}
 
 Besides those traditional dimensionless numbers, this system introduces $\Upsilon$, relating the large-scale kinetic energy to the energy dissipated by the unresolved component:
\begin{equation}
\Upsilon
= \frac{UL}{A_u }
= \frac{U^2}{A_u / T}.
\end{equation}


\section{QG model under moderate uncertainty}
\label{appendix QG model under moderate uncertainty}

Hereafter, we consider the QG approximation ($\Ros  \ll 1$ and $\Bu \sim 1$), for $\Upsilon \sim 1$. We focus on solutions of  the Boussinesq model with Rossby number going to zero. To derive the evolution equations corresponding to this limit, the solution of the non-dimentional Boussinesq model (Appendix \ref{appendix Nondimensionalized Boussinesq equations}) is developed as a power series of the Rossby number:
\begin{eqnarray}
\begin{pmatrix}
\w \\ b \\ \phi
\end{pmatrix}
= \sum_{k=0}^{\infty} 
 \Ros^k 
\begin{pmatrix}
\w_k \\ b_k \\ \phi_k
\end{pmatrix}
.
\label{Rossby Taylor series}
\end{eqnarray}
According to the horizontal momentum equation \eqref{sto-Boussinesq-buo:Momentum equations}, the scaling of the pressure still corresponds to the usual geostrophic balance. This sets the Euler number as:
\begin{eqnarray}
\Eu \sim \frac 1 \Ros.
\end{eqnarray}
For the ocean, the aspect ratio, $D\defi H /L$, is small and $D^2\ll1$. As a consequence,
\begin{eqnarray}
\frac{D^2 }{\Eu } \sim D^2 \Ros  \ll D^2  \ll 1
\text{ and }
\frac{D^2 }{\Eu \Upsilon} \sim \frac{D^2 \Ros }{ \Upsilon} \leqslant D^2  \ll 1.
\end{eqnarray}
Therefore, the inertial and diffusion terms are negligible in the vertical momentum equation. The hydrostatic assumption is still valid. This leads to the classical QG scaling of the buoyancy equation:
\begin{equation}
\Gamma \sim  \Eu \sim \frac 1 \Ros
\text{ and }
\frac 1 { \Fr^2 \Gamma} \sim \frac \Ros {\Fr^2} = \frac \Bu \Ros.
\end{equation}
In the following, the subscript $H$ is omitted for the differential operators Del, $\nab$, and Laplacian, $\Delta$. They all represent $2$D operators. Only keeping terms of order $0$ and $1$, we get the following system:

\begin{align}
&\hspace*{-0.5cm}\text{\em Momentum equations} \nonumber\\
&
\label{scaled Boussinesq QG horiz momentum}
\Ros \left( 
\dif_t\mbs u 
+ \left(\mbs u \bcdot \nab \right) \mbs u \dif t
+  \frac{1}{\Upsilon^{1/2} }\left( \bsigma_H \dBt  \bcdot \nab \right) \mbs u
-  \frac{1}{2 \Upsilon} \sum_{i,j\in H}  \partial_{ij}^2\bigl (a_{ij} \mbs u\bigr)  \dif t
+  O \left( \frac \Ros {\Upsilon \Bu} \right) 
\right)
\nonumber \\ & \hspace{0.8cm}
+\left( f_0 + \Ros \beta y \right)  \mbs k  \times 
\left( \mbs u \dif t +  \frac{1}{\Upsilon^{1/2} }\bsigma_H \dBt \right)
=
- \;\nab_H \left( \phi' \dif t + \frac{1}{\Upsilon^{1/2} } \dif_t \phi_\sigma \right) 
  ,\\
&
 b \; \dif t 
+  O \left(  \Ros D^2 \right) 
 = \;\partial_z\left( \phi' \dif t + \frac{1}{\Upsilon^{1/2} } \dif_t \phi_\sigma \right) , 
\label{scaled Boussinesq QG vert momentum}
 \\
&\hspace*{-0.5cm}\text{\em Buoyancy equation} \nonumber\\
&
\label{scaled Boussinesq QG buoy}
\frac{ \Ros}{\Bu} \left( 
{\dif_t b} 
+ \nab b \bcdot \; \left( \mbs u \dif t +   \frac{1}{\Upsilon^{1/2} }(\bsigma d\B_{t})_H \right)  
+ \partial_z b \ w \dif t
-  \frac{1}{2 \Upsilon} \sum_{i,j\in H}  \partial_{ij}^2\left(a_{ij} b \right)  \dif t
\right)
\nonumber \\ &
 \hspace{1.5cm}
+   w\dif t 
- \frac 1{ \Upsilon}  \frac \Ros \Bu \left ( \nab \bcdot a_{Hz}\right) \transp \dif t 
+  \frac \Ros \Bu \frac{1}{\Upsilon^{1/2}} (\bsigma \dif \B_t)_z
+ O \left( \frac {\Ros^2} {\Upsilon\Bu^2} \right) 
=0,
 \\
&\hspace*{-0.5cm}\text{\em Incompressibility} \nonumber\\
&
 \nab \bcdot \mbs u+ \partial_z w =0,
 \label{QG deriv incomp1}\\
& 
 \nab {\bcdot}\bigl(\bsigma \dif \B_t\bigr)_H  
 + \frac \Ros \Bu \partial_z\bigl(\bsigma \dif \B_t\bigr)_z
 =0,
 \label{QG deriv incomp2}\\
& 
  \nab \bcdot \left( \nab \bcdot \mbs a_H \right) \transp
 + 2 \frac \Ros \Bu \nab \bcdot \partial_z \mbs a_{Hz}
 + O \left( \left( \frac \Ros \Bu \right)^2 \right)
 =0.
 \label{QG deriv incomp3}
\end{align}

The thermodynamic equation \eqref{scaled Boussinesq QG buoy} at $0$ order leads to :
\bea 
w_0 = 0,
\eea 
and then, by the large-scale incompressibility equation \eqref{QG deriv incomp1}, the $0$-order horizontal velocity is divergence-free. Following the scaling assumption, the horizontal small-scale velocity is also divergence-free \eqref{QG deriv incomp2}. The horizontal momentum equation \eqref{scaled Boussinesq QG horiz momentum}  at the $0$-th order leads to:
\begin{eqnarray}
\mbs u_0  = 
\frac 1 f_0
 \nab^{\bot} \phi_0' 
\text{ and }
(\bsigma \dif \B_t)_{H}  = 
\frac 1 f_0
 \nab^{\bot} \dif_t \phi_{\sigma}
,
\end{eqnarray}
where time-correlated and time-uncorrelated components have been separated by the mean of uniqueness of the semi-martingale decomposition \citep{Kunita}. Being divergent-free, both components can be expressed with two stream functions $\psi_0$ and $\dif_t \psi_{\sigma}$:
\begin{eqnarray}
\mbs u_0  =  \nab^{\bot} \psi_0
\text{ and }
(\bsigma \dif \B_t)_{H}  =  \nab^{\bot} \dif_t \psi_{\sigma },
\label{random-pressure-H}
\end{eqnarray}
exactly corresponding to the dimensionless pressure terms:
\begin{eqnarray}
\psi_0 = 
\frac 1 f_0
  \phi_0' 
\text{ and }
\dif_t \psi_{\sigma} =
\frac 1 f_0
 \dif_t \phi_{\sigma}
.
\end{eqnarray}
Deriving these equations along $z$ and introducing the hydrostatic equilibrium \eqref{scaled Boussinesq QG vert momentum} -- decomposed between correlated and uncorrelated components -- yields the classical thermal wind balance at large-scale for the $0$-th order terms. The buoyancy variable does not involve any  white noise term, and the small-scale random velocity is thus almost constant along $z$, as
\begin{eqnarray}
\partial_z \mbs u_0  =
\frac 1 f_0
 \nab^{\bot} b_0
\text{ and }
\partial_z (\bsigma \dif \B_t)_{H}  = O \left( \Ros D^2 \right )
.
\label{thermal wind QG sto}
\end{eqnarray}
Accordingly the variance tensor scales as:
\begin{eqnarray}
\forall i,j \in H, \ \partial_z a_{ij} =  O \left( \Ros^2 D^4 \right ),
\end{eqnarray}
which is negligible in all equations, and the uncertain random field solely depends on the horizontal coordinates. Since $\Ros/ \Bu \sim \Ros$, the $1$-st order term of the buoyancy equation must be kept to describe the evolution of $b_0$:
\begin{eqnarray}
\frac 1 \Bu \Dt^H_{0t} b_0 
+   w_{1} \dif t 
- \frac 1{ \Upsilon}  \left ( \nab \bcdot a_{Hz}\right) \transp  \dif t 
+ \frac{1}{\Upsilon^{1/2}} (\bsigma \dif \B_t)_{z}
=0,
\end{eqnarray}
where, for all functions $h$,
\begin{eqnarray}
\Dt^H_{0t} h \defi 
{\dif_t h} 
+ \nab h \bcdot \; \left( \mbs u_0 \dif t +   \frac{1}{\Upsilon^{1/2} }(\bsigma \dif \B_{t})_{H} \right)  
-  \frac{1}{2 \Upsilon} \sum_{i,j\in H}  \partial_{ij}^2 \left(a_{ij} h \right) \dif t.
\end{eqnarray}
Taking the derivative along $z$ leads to:
\begin{multline}
\frac 1 \Bu \left(
\Dt^H_{0t} \partial_z b_0 
+ \nab b_0 \bcdot \; \partial_z \left( \mbs u_0 \dif t +   \frac{1}{\Upsilon^{1/2} }(\bsigma d\B_{t})_{H} \right)  
-  \frac{1}{2 \Upsilon} \sum_{i,j\in H}  \partial_{ij}^2\left( \partial_z a_{ij} b_0 \right)  \dif t \right)
\\
 + \partial_z w_{1} \dif t 
- \frac 1{ \Upsilon}  \left ( \nab \bcdot \partial_z a_{Hz}\right) \transp  \dif t 
 + \frac{1}{\Upsilon^{1/2}} \partial_z (\bsigma \dif \B_t)_{z}
=0.
\end{multline}
The introduction of the thermal wind equations (\ref{thermal wind QG sto}) and incompressibility conditions (\ref{QG deriv incomp1}-\ref{QG deriv incomp3}) helps simplifying this equation as:
\begin{multline}
\frac 1 \Bu  \Dt^H_{0t} \partial_z b_0 
- \dv  \mbs u_{1} \dif t
+ \left ( \frac \Ros \Bu \right)^{\!-1} \!\!\frac 1 {  \Upsilon} \nab \bcdot \left( \nab \bcdot \mbs a_H \right) \transp  \dif t
- \left ( \frac \Ros \Bu \right)^{\!-1} \!\!\frac{1}{\Upsilon^{1/2}} \dv (\bsigma \dif \B_t)_{H}
=0.
\label{buyancy QG deriv}
\end{multline}
Note the factor $\left ( \frac \Ros \Bu \right)^{\!-1} $ appears. It comes from the incompressible conditions \eqref{QG deriv incomp2} and \eqref{QG deriv incomp3}, leading $ \dv (\bsigma \dif \B_t)_{H}$ and $\nab \bcdot \left( \nab \bcdot \mbs a_H \right) \transp \dif t$ to both scale as  $\frac \Ros \Bu $. The hydrostatic balance at $0$-order links the buoyancy to the pressure, and then to the stream function
\begin{eqnarray}
 \partial_z b_0 =  \partial^2_z \phi_0 = 
 f_0
 \partial^2_z \psi_0.
\label{link buyancy QG deriv}
\end{eqnarray}
The $1$-st order term of the vertical velocity is not known. Yet, the system can be closed using the vorticity equation at order $1$:

\begin{multline}
\nab^{\bot} \!\bcdot\! \left( 
\Dt^H_{0t}  \mbs u_0 
\right)
+   
f_0 \left(
\dv \mbs u_1 
+
\left ( \frac \Ros \Bu \right)^{-1}
 \dv (\bsigma \dif \B_t)_H
\right)
+  \bnabla (\beta y)\! \bcdot 
\left ( \mbs u_0 \dif t + (\bsigma \dif \B_t)_H
\right ) 
=   
0 ,
\end{multline}
where the divergence terms come from the constant Coriolis term.

Again, factors $\left ( \frac \Ros \Bu \right)^{-1}$ compensate the order of magnitude of $ \dv (\bsigma \dif \B_t)_{H}$ and $\nab \bcdot \left( \nab \bcdot \mbs a_H \right) \transp \dif t$. Then,
\begin{multline}
\Dt^H_{0t} \left( \Delta \psi_0
\right)
+
f_0
 \dv u_1\dif t
+ \frac 1 {\Upsilon^{1/2}} \left ( \frac \Ros \Bu \right)^{-1}
  f_0
 \dv (\bsigma \dif \B_t)_H
  +  \beta \left( v_0 \dif t + \frac 1 {\Upsilon^{1/2}}(\bsigma \dif \B_t)_y \right)
  \\
  + \frac 1 {\Upsilon^{1/2}} \tr \left( \nab^{\bot}  (\bsigma \dif \B_t)_H \transp  \nab \mbs u_0 \transp
   \right )
   -  \frac{1}{2 \Upsilon} \sum_{i,j\in H}  \partial_{ij}^2 \left( \nab^{\bot} a_{ij} \bcdot  \mbs u_0 \right) \dif t
   =0.
\end{multline}
Using (\ref{buyancy QG deriv}) and (\ref{link buyancy QG deriv}), we get:
\begin{multline}
\Dt^H_{0t} \left( \Delta \psi_0 + f_0 +\beta y + 
\frac {f_0^2}{\Bu}
 \partial_z^2 \psi_0
\right)
=-  \dv a_{Hy} \beta \dif t
- 
f_0
 \left ( \frac \Ros \Bu \right)^{-1} \frac 1 { 2 \Upsilon} \nab \bcdot \left( \nab \bcdot \mbs a_H \right) \transp  \dif t
  \\
  - \tr \left( \nab^{\bot}  (\bsigma \dif \B_t)_H \transp  \nab \mbs u_0 \transp
   \right )
   +  \frac{1}{2 \Upsilon} \sum_{i,j\in H}  \partial_{ij}^2 \left( \nab^{\bot} a_{ij} \bcdot  \mbs u_0 \right)     \dif t.
\end{multline}
We recall that coefficients $\left ( \frac \Ros \Bu \right)^{-1}$ are still present since
\bea 
\dv (\bsigma \dif \B_t)_{H} \sim \nab \bcdot \left( \nab \bcdot \mbs a_H \right) \transp \dif t \sim \left ( \frac \Ros \Bu \right).
\eea
If we rewrite the equation with dimensional quantities, the evolution equation for $u_0 = \lim_{\Ros \to 0} u $ is obtained (dropping the index $0$ for clarity):
\begin{multline}
\label{appendix first conservation PV moderate uncertainty}
\Dt^H_{t} Q
= 
- \tr \left( \nab^{\bot}  (\bsigma \dif \B_t)_H \transp  \nab \mbs u \transp
   \right )
+  \frac{1}{2} \sum_{i,j\in H}  \partial_{ij}^2 \left( \nab^{\bot} a_{ij} \bcdot  \mbs u \right)     \dif t
-  \frac 1 2 \nab \bcdot \left( \nab \bcdot (\mbs a_H f) \right) \transp  \dif t 
   ,
\end{multline}
where $Q$ is the QG potential vorticity:
\begin{eqnarray}
Q \defi  \Delta \psi +f+ \left( \frac {f_0} N \right )^2 \partial_z^2 \psi.
\end{eqnarray}
Note, \eqref{random-pressure-H} provides the geostrophic balance for the small-scale velocity component. To express the material derivative of $Q$, the noise term is expanded:
\bea 
- \tr \left( \nab^{\bot}  (\bsigma \dif \B_t)_H \transp  \nab \mbs u \transp
   \right )
   =
 - \sum_{k,j \in H}
 \partial_{kj}^2 \psi
\partial_{k}
\bsigma_{j\bullet} \dBt
.
\eea
According to \cite{resseguier2016geo1}, the difference between the material derivative, $D_t Q$, and the stochastic transport operator $\Dt_t Q$, is a function of the time-uncorrelated forcing:
\bea 
\label{link DD and material deriv}
\left\{
\begin{array}{r c l}
{ \Dt}_t Q &= &
f_1 \dif t +  \mbs h_1 \transp \dif\B_t, 
\label{Dt-eq1}\\
D_t Q &=& 
f_2 \dif t +  \mbs h_2 \transp \dif\B_t,
  \end{array}
  \right.
\Longleftrightarrow 
\left\{
\begin{array}{r c l}
f_2 &=& 
f_1  
+ \tr\bigl(\left ( \bsigma \transp \bnabla \right )  \mbs h_1 \transp \bigr),\\
\mbs h_1 &=& 
\mbs h_2 .
  \end{array}
    \right.
  \eea 
The expression of $h_1$ is given by equation \eqref{appendix first conservation PV moderate uncertainty} and the above formulas give:

\bea 
D_t Q - \Dt_t Q
&=&
 \sum_{i \in H}
\bsigma_{i \bullet}
\partial_i \left( 
- \sum_{j,k \in H}
 \partial_{k}
\bsigma_{j\bullet}
\partial_{kj}^2 \psi
 \right) \transp
 ,\\
&=&
- \sum_{i,j,k \in H}
\left(
\bsigma_{i \bullet}
 \partial_{ik}^2
\bsigma_{j\bullet} \transp
\partial_{kj}^2 \psi
+
\bsigma_{i \bullet}
 \partial_{k}
\bsigma_{j\bullet} \transp
\partial_{ijk}^3 \psi
\right)
.
 \label{appendix expand 1}
\eea
With the use of the small-scale incompressibility, we obtain:
 \begin{multline}
 \label{appendix expand 2}
  \frac{1}{2} \sum_{i,j\in H}  \partial_{ij}^2 \left( \nab^{\bot} a_{ij} \bcdot  \mbs u \right)
  =
  \\
 \sum_{i,j,k\in H}
  \left( 
  \partial_j \bsigma_{i\bullet} 
 \partial_{ik}^2 \bsigma_{j\bullet} \transp
 \partial_k \psi
+
 \partial_j \bsigma_{i\bullet}
\partial_{k} \bsigma_{j\bullet} \transp
\partial_{ik}^2 \psi
+
 \bsigma_{i\bullet}
\partial_{ik}^2 \bsigma_{j\bullet} \transp
\partial_{jk}^2 \psi
+
 \bsigma_{i\bullet}
\partial_{k} \bsigma_{j\bullet} \transp
\partial_{ijk}^3 \psi
  \right).
 \end{multline}
From \eqref{appendix expand 1} and \eqref{appendix expand 2}, it yields:
\begin{multline}
D_t Q
-
\left(
\Dt_t Q 
- \frac{1}{2} \sum_{i,j\in H}  \partial_{ij}^2 \left( \nab^{\bot} a_{ij} \bcdot  \mbs u \right)
\right)
= 
\sum_{i,j,k \in H}
\left(
\partial_j \bsigma_{i\bullet} 
\partial_{ik}^2 \bsigma_{j\bullet} \transp
\partial_k \psi
+
\partial_j \bsigma_{i\bullet} 
\partial_{k} \bsigma_{j\bullet} \transp
\partial_{ik}^2 \psi
\right).
\end{multline}

Denoting, $\alpha$, the following matrix
\begin{eqnarray}
\alpha_{ij} 
\defi 
\sum_{k \in H}\partial_k \bsigma_{i\bullet} \partial_j \bsigma_{k \bullet} \transp
=
\sum_{k \in H}\partial_k (  \bsigma_{i\bullet} \partial_j \bsigma_{k \bullet} \transp ) ,
\end{eqnarray}
we have
\bea 
  \nab \bcdot ( \alpha \nab \psi )
  &=&
\sum_{i,j,k \in H}
\left(
\partial_j \bsigma_{i\bullet} 
\partial_{ik}^2 \bsigma_{j\bullet} \transp
\partial_k \psi
+
\partial_j \bsigma_{i\bullet} 
\partial_{k} \bsigma_{j\bullet} \transp
\partial_{ik}^2 \psi
\right),
\\
&=&
D_t Q
-
\left(
\Dt_t Q 
- \frac{1}{2} \sum_{i,j\in H}  \partial_{ij}^2 \left( \nab^{\bot} a_{ij} \bcdot  \mbs u \right)
\right),
 \eea
and the material derivative of the PV finally reads:
\bea 
D^H_t Q
= 
  \nab \bcdot ( \alpha \nab \psi )    \dif t
-  \frac 1 2 \nab \bcdot \left( \nab \bcdot (\mbs a_H f) \right) \transp  \dif t
- \tr \left [
 \nab^{\bot}  (\bsigma \dif \B_t)_H \transp  \nab \mbs u \transp
   \right ].
   \label{transport of PV appendix}
 \eea
To note, the transpose of the matrix $\alpha$ has a compact expression:
\begin{eqnarray}
\alpha \transp
=
\sum_p
\left( \nab \bsigma_{Hp}\transp \right)^2.
\end{eqnarray}
To better assess the role of the random source term (the last term of \eqref{transport of PV appendix}), it is decomposed in terms of symmetric and anti-symmetric parts of the small-scale/large-scale deformation tensors. Let us denote $\mbs S$ and $ \mbs S_{ \sigma \dif B_t}$ the symmetric parts of $\nab \mbs u \transp$ and $\nab  (\bsigma \dif \B_t)_H \transp$, respectively. Associated with divergence-free velocities, these symmetric parts, so-called strain rate tensors, have zero trace. Terms $-\frac 12 \omega \mbs J$ and $-\frac 12\omega_{ \sigma \dif B_t} \mbs J$ will stand for the anti-symmetric parts, where $
\mbs J = 
\begin{pmatrix}
0 & -1 \\
1 & 0
\end{pmatrix}$ is the $\frac {\pi} 2$ rotation. The factors $\omega$ and $\omega_{ \sigma \dif B_t}$ are the large-scale and the small-scale components of the vorticity, respectively. Using $ \mbs J \mbs J = - \id $ and $\tr [ \mbs M \mbs N]=tr[ \mbs N \mbs M]$ yields:
\bea
- \tr \left [
 \nab^{\bot}  (\bsigma \dif \B_t)_H \transp  \nab \mbs u \transp
   \right ]
   &=&
- \tr \left [ \mbs J
    \left( 
     \mbs S_{ \sigma \dif B_t} -\frac 12 \omega_{ \sigma \dif B_t} \mbs J
    \right )
 \left( 
 \mbs S -\frac 12 \omega \mbs J 
    \right )
    \right ], \\
   &=&
    \nonumber
- \tr \left [ 
 \mbs S \mbs J
  \mbs S_{ \sigma \dif B_t}
    \right ]
 -\frac 12 \omega_{ \sigma \dif B_t}
 \underbrace{
   \tr \left [ \mbs S \right ]
   }_{=0} \\
   & &
 -\frac 12 \ \omega  
 \underbrace{
   \tr \left [  \mbs S_{ \sigma \dif B_t} \right ]
   }_{=0}
 +\frac 14 \ \omega \ \omega_{ \sigma \dif B_t}
 \underbrace{
   \tr \left [ \mbs J \right ]
   }_{=0}, \\
   &=&
- \tr \left [ 
 \mbs S \mbs J
  \mbs S_{ \sigma \dif B_t}
    \right ].
\eea
This term thus only depends on the strain rate tensors of $\mbs u$ and $(\bsigma \dif \B_t)_H$. The PV transport can thus be rewritten as:
\bea
D^H_t Q
= 
  \nab \bcdot ( \alpha \nab \psi )    \dif t
-  \frac 1 2 \nab \bcdot \left( \nab \bcdot (\mbs a_H f) \right) \transp  \dif t
- \tr \left [ 
 \mbs S \mbs J
 \mbs S_{ \sigma \dif B_t}
 \right ].
 \eea
The noise term can be further expressed using the stable directions of the flows defined by $\mbs u$ and $(\bsigma \dif \B_t)_H$, respectively. In the following, we will omit writing the $\dBt$ factor. The two strain rate tensors are decomposed in orthogonal basis:
\bea 
\mbs S 
= \mbs V \mbs \Xi \mbs V \transp
= \sum_{p=1}^2 \Xi_{pp} {\mbs V}_{\bullet p} {\mbs V}_{\bullet p} \transp
\text{ and }
 \mbs S_{ \sigma \dif B_t}
= \mbs W \mbs \Lambda \mbs W \transp,
\eea
where $\mbs V_{\bullet p} \transp \mbs V_{\bullet q} = \mbs W_{\bullet p} \transp \mbs W_{\bullet q} = \delta_{pq}$, $\Xi_{11} = - \Xi_{22} <0$ and $\Lambda_{11} = - \Lambda_{22} <0$.
\bea
- \tr \left [ 
 \mbs S \mbs J
 \mbs S_{ \sigma \dif B_t}
    \right ]
    &=&
  -  \sum_{p,q=1}^2 \Xi_{pp} \Lambda_{qq}
\tr \left [ 
{\mbs V}_{\bullet p} {\mbs V}_{\bullet p} \transp \mbs J
{\mbs W}_{\bullet q} {\mbs W}_{\bullet q} \transp
    \right ],\\
    &=&
  -  \sum_{p,q=1}^2 \Xi_{pp} \Lambda_{qq}
\left (
{\mbs V}_{\bullet p} \transp {\mbs W}_{\bullet q}
    \right )
\left (
{\mbs V}_{\bullet p} \transp \mbs J {\mbs W}_{\bullet q}
    \right ),\\
    &=&
  -  \sum_{p,q=1}^2 \Xi_{pp} \Lambda_{qq} \ 
\cos(\theta_{pq}) \cos \left(\theta_{pq} + \frac \pi 2 \right),\\
    &=&
     \frac 12 \sum_{p,q=1}^2 \Xi_{pp} \Lambda_{qq} \ 
\sin(2 \theta_{pq} ),
\eea
where $\theta_{pq}\defi \widehat{ \left(\mbs V_{\bullet p},\mbs W_{\bullet q} \right)}$ is the angle between ${\mbs V}_{\bullet p}$ and ${\mbs W}_{\bullet q}$. Using the relations between the eigenvalues and the orthogonality of the eigenvectors, it finally comes:
\bea
- \tr \left [ 
 \mbs S \mbs J
 \mbs S_{ \sigma \dif B_t}
    \right ]
    &=&
    \nonumber
  \frac 12 \Xi_{11} \Lambda_{11} 
 \left(
 \sin \left( 2\theta_{11} \right) 
 - \sin \left(2\left(\theta_{11} - \frac{\pi} 2\right)\right) 
 - \sin\left(2\left(\theta_{11} + \frac{\pi} 2\right)\right) 
 + \sin\left(2\theta_{11} \right)
 \right),\\
    &=&
  2 \underbrace{ \Xi_{11} \Lambda_{11} }_{>0}
 \sin \left( 2 \theta_{11} \right) .
\eea


\end{document}